\PassOptionsToPackage{dvipsnames}{xcolor}
\documentclass[aps,prapplied,reprint,a4paper,floatfix,amsmath,amssymb,amsfonts,noshowpacs,longbibliography]{revtex4-2}
\usepackage{newtxtext,newtxmath}
\usepackage[T1]{fontenc}
\usepackage[utf8]{inputenx}
\input{ix-utf8enc.dfu}
\usepackage{graphicx}
\usepackage{sidecap}
\sidecaptionvpos{figure}{t}
\usepackage[dvipsnames]{xcolor}
\usepackage{soul} 
\usepackage{xspace}
\usepackage{siunitx}
\usepackage{url}
\usepackage{floatrow}
\usepackage[
,textwidth=17.5cm
,textheight=23.7cm
,verbose
,pdftex
]{geometry}
\usepackage{xspace}

\usepackage[pdftex]{hyperref}
\hypersetup{
	pdftitle={Electrical properties of ScN layers grown on GaN(0001) by plasma-assisted molecular beam epitaxy},
	colorlinks,
	citecolor=blue,
	linkcolor=blue,
	urlcolor=blue
}

\newcommand{\munit}{\,cm$^{2}$V$^{-1}$s$^{-1}$\xspace}
\newcommand{\nunit}{\,$\text{cm}^{-3}$\xspace}
\newcommand{\K}{\,K}
\newcommand{\mH}{$\mu_\text{H}$}
\newcommand{\nH}{$n_\text{H}$}
\newcommand{\mCB}{$\mu_\text{CB}$}
\newcommand{\nCB}{$n_\text{CB}$}

\newcommand{\mHs}{$\mu_\text{H}$\xspace}
\newcommand{\nHs}{$n_\text{H}$\xspace}
\newcommand{\mCBs}{$\mu_\text{CB}$\xspace}
\newcommand{\nCBs}{$n_\text{CB}$\xspace}
\newcommand{\mIBs}{$\mu_\text{IB}$\xspace}
\newcommand{\nIBs}{$n_\text{IB}$\xspace}

\newcommand{\PAMBE}{plasma-assisted molecular beam epitaxy\xspace}
\newcommand{\sap}{Al$_2$O$_3$}
\newcommand{\x}{\times}

\begin{document}

\title{Electrical properties of ScN(111) layers grown on GaN(0001)\\ by plasma-assisted molecular beam epitaxy}

\author{Duc V. Dinh}
\email[Electronic email: ]{dinh@pdi-berlin.de}
\author{Oliver Brandt}
\affiliation{Paul-Drude-Institut für Festkörperelektronik, Leibniz-Institut im Forschungsverbund Berlin e.V., Hausvogteiplatz 5--7, 10117 Berlin, Germany.}

\date{\today}

\begin{abstract}
We investigate the electrical properties of nominally undoped, 10--40-nm-thick ScN(111) layers grown on nearly lattice-matched GaN:Fe/\sap(0001) templates by plasma-assisted molecular beam epitaxy. Hall-effect measurements yield electron concentrations of 0.7--3.1$\times 10^{19}$\nunit and mobilities of 50--160\munit at room temperature. The temperature-dependent (4--360\K) conductivity exhibits two distinct regimes, suggesting two-band conduction in an impurity band and the conduction band. Assuming a single shallow donor in ScN and employing the standard two-band conduction model, we extract the carrier density and mobility in these bands. The results reveal nondegenerate characteristics for the 40-nm-thick layer, while the thinner layers are weakly degenerate. For the nondegenerate layer, the donor ionization energy amounts to approximately 12\,meV. The electron mobility of the layers is limited by ionized impurity scattering and phonon scattering at low and high temperatures, respectively. Fits with an expression for optical phonon scattering developed for weakly degenerate semiconductors return an effective phonon energy of $(61 \pm 5)$\,meV, in between the energies of the longitudinal optical ($\approx 84$\,meV) and transverse optical ($\approx 45$\,meV) phonon modes in ScN.
\end{abstract}

                  
\maketitle


\section{Introduction}

ScN is an emerging group-III$_\text{B}$ transition-metal nitride semiconductor crystallizing in the rocksalt structure. The electronic band structure of ScN is characterized by an indirect gap at about 0.9\,eV between the $X$ and $\Gammaup$ points \cite{Al-Brithen2004Jul,Saha2010Feb,Deng2015Jan,Cetnar2018Nov,Mu2021Aug}, a direct gap at about 2\,eV at the $X$ point \cite{Dismukes1972May,Smith2001Aug, Deng2015Jan,Al-Brithen2004Jul,Moram2008Oct,Saha2010Feb,Saha2013Aug,Oshima2014Apr,Lupina2015Nov,Cetnar2018Nov,Rao2020Apr,Dinh2023Sep}, and higher direct gaps at the $\Gammaup$ point of about 3.89\,eV \cite{Deng2015Jan,Cetnar2018Nov,Mu2021Aug,Dinh2023Sep}, 5.33\,eV \cite{Deng2015Jan,Dinh2023Sep} and 6.95\,eV \cite{Deng2015Jan,Dinh2023Sep}. 

ScN has attracted significant interest in recent years for its potential thermoelectric \cite{Kerdsongpanya2011Dec,Burmistrova2013Apr,Rao2020Apr} and infrared optoelectronic applications \cite{Maurya2022Jul}. Due to its close lattice match to GaN ($\Delta a/a \approx 0.1\%$) and a large polarization discontinuity at the ScN(111)/GaN\{0001\} interface, ScN is also considered a very promising material for high density two-dimensional electron and hole gases \cite{Adamski2019Dec}. Additionally, ScN has attracted much interest in the form of alloys with the conventional group-III$_\text{A}$ nitrides, i.\,e., GaN, AlN, and InN. In particular, the ternary alloy (Sc,Al)N holds great potential for applications in surface acoustic wave devices \cite{Akiyama2009Feb,Hashimoto2013Mar}, field-effect transistors \cite{Hardy2017Apr,Wang2021Aug,Dinh2023Apr}, and as novel ferroelectric material \cite{Wang2022Jul, Wang2023Jan}.

Regarding its electrical properties, nominally undoped ScN is invariably $n$-type with electron densities \nHs ranging from $10^{19}$ to $10^{21}$\nunit at room temperature \cite{Dismukes1972May,Moram2008Oct,Burmistrova2013Apr,Casamento2019Oct,Ohgaki2013Sep,Deng2015Jan,Cetnar2018Nov,Rao2020Apr,Al-Atabi2020Mar,Dinh2023Sep}. In sputter-deposited material, the high electron density often stems from contamination of the target with F \cite{Deng2015Jan,Cetnar2018Nov}. In material synthesized by hydride vapor phase or molecular beam epitaxy, the background doping is attributed to O \cite{Moram2008Oct,Casamento2019Oct,Dinh2023Sep,Cetnar2018Nov}. The lowest \nHs of $3.7 \x 10^{18}$\nunit and simultaneously highest mobility (\mH) of 284\munit at room temperature have been reported for 40-$\muup$m-thick ScN layers grown on Al$_2$O$_3$ substrates by hydride vapor phase epitaxy, and were attributed to the low impurity concentrations in those layers (in particular, $[\text{O}] < 1 \x 10^{18}$\nunit) \cite{Oshima2014Apr}.

Several groups have investigated the electrical transport in unintentionally doped $n$-type ScN using temperature-dependent Hall-effect measurements \cite{Al-Atabi2020Mar,Cetnar2018Nov,Casamento2019Oct,Ohgaki2013Sep,Rao2020Apr}. With the commonly observed background doping levels ranging from $10^{20}$--$10^{21}$\nunit, ScN is highly degenerate, i.\,e., \nHs does not depend on temperature \cite{Cetnar2018Nov,Casamento2019Oct,Al-Atabi2020Mar}. Additionally, \mHs of those layers has been found to be independent of temperature \cite{Casamento2019Oct} or to only weakly decrease with increasing temperature \cite{Cetnar2018Nov,Al-Atabi2020Mar}. The low mobility observed in the former case has been suggested to possibly arise from boundary scattering in the twinned ScN(111) films \cite{Casamento2019Oct}. 

The decrease of mobility for increasing temperatures has been analyzed quantitatively in only a few studies \cite{Dismukes1972May,Saha2017Jun}. The temperature dependence was found to follow a $T^{-\alpha}$ dependence with $\alpha$ being close to the exponent 3/2 expected for acoustic phonon scattering. Other authors, not attempting a quantitative analysis, have attributed this decrease of mobility to dislocation scattering \cite{Rao2020Apr} or optical phonon scattering \cite{Cetnar2018Nov,Al-Atabi2020Mar}. In fact, theory shows that optical phonon scattering should be the dominant mechanism rather than acoustic phonon scattering \cite{Mu2021Aug}.

In this article, we investigate the electrical properties of 10--40-nm thick ScN(111) layers grown on nearly lattice-matched GaN(0001) templates using \PAMBE (PAMBE). The electrical properties of the layers are investigated in detail by temperature-dependent Hall-effect measurements, yielding comparatively low electron concentrations of 0.7--3.1$\times 10^{19}$\nunit and comparatively high mobilities of 50--160\munit at room temperature. At the same time, the data reveal the presence of parallel conduction originating from both impurity and conduction bands. The electron density and mobility in the conduction band are extracted using a standard two-band conduction model and analyzed with models valid for weakly degenerate semiconductors. The temperature dependence of the electron mobility in the conduction band is limited by ionized-impurity scattering at low and optical phonon scattering at high temperatures. Peak values up to 280\munit are observed at temperatures of 140--180\,K, which are among the highest mobilities ever reported for ScN.

\section{Experimental}

ScN layers are grown by PAMBE on semi-insulating GaN:Fe/Al$_2$O$_3$(0001) templates fabricated by metal-organic vapor phase epitaxy. Before being loaded into the ultrahigh vacuum environment, the GaN templates are etched in an HCl solution to remove the surface oxide as well as surface contaminants, and then rinsed with de-ionized water and finally blown dry with a nitrogen gun. Afterwards, the templates are outgassed for 2 hours at 500\,°C in a load-lock chamber attached to the MBE system. The MBE growth chamber is equipped with high-temperature effusion cells to provide the group III metals, including 7N pure Ga and 5N pure Sc (note that these purity levels refer to trace metals, but not to elements such as N, C, or O). A Veeco UNI-Bulb radio-frequency plasma source is used for the supply of active nitrogen (N$^*$). We use 6N N$_2$ gas as precursor which is further purified by a getter filter. The N$^*$ flux is calculated from the thickness of a GaN layer grown under Ga-rich conditions and thus with a growth rate limited by the N$^*$ flux. Prior to ScN growth, a 100-nm-thick undoped GaN buffer layer is grown at 700\,°C under Ga-bilayer conditions. Subsequently, ScN layers are grown at the same temperature under N$^*$-rich conditions with thicknesses ($d_\text{ScN}$) ranging from 10 to 40\,nm.

The layers exhibit excellent structural and morphological properties as evidenced in Figs.\,S1 and S2 in the supplementary material. For surface chemical analysis, x-ray photoelectron spectroscopy (XPS) measurements have been performed using a Scienta Omicron\texttrademark\ system equipped with an Al anode (Al\,K$\alpha$: $h\nu = 1486$\,eV) under ultra-high vacuum (UHV) conditions. XPS surveys were measured before and after Ar$^+$ sputtering with the intention to remove surface oxides and contaminants (see Fig.\,S3 in the supplementary material). Information regarding impurities in the Sc source, as provided by the commercial vendor, is presented in Fig.\,S4 of the supplementary material. Raman spectra of all the layers are recorded at room temperature using a Horiba LabRAM HR Evolution\texttrademark\ Raman microscope using a 473-nm diode-pumped solid state laser for excitation (see Fig.\,S5 in the supplementary material).

To investigate the electrical properties of the layers, we carry out Hall-effect measurements in the van der Pauw configuration at 4--360\,K. A magnetic field of 0.7\,T is applied for these measurements. The room-temperature carrier density (\nH) and mobility (\mH) are shown in Table.~\ref{tab:table}, as well as the values for the conduction band (CB) after accounting for parallel impurity band (IB) conduction. For comparison, temperature-dependent Hall-effect measurements are also performed on a ScN/Al$_2$O$_3$(0001) reference layer \cite{Dinh2023Sep}.

\begin{figure}
\includegraphics[width=\columnwidth]{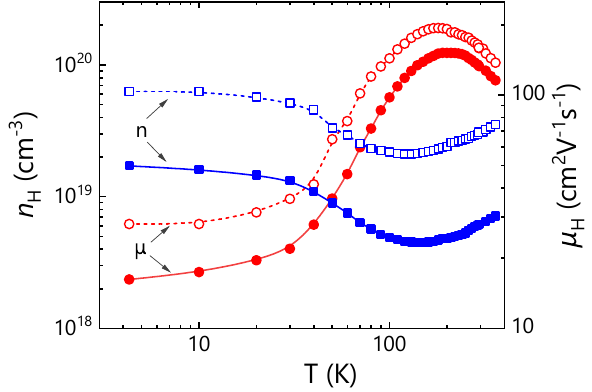}
	\caption{Temperature dependence of $n_\text{H}$ and $\mu_\text{H}$ for the ScN layers with $d_\text{ScN} = 10$\,nm ($\square, \circ$) and $d_\text{ScN} = 40$\,nm ($\blacksquare, \bullet$).}
	\label{fig:Hall}
\end{figure}

\begin{table}[b]
\caption{Overview of the electrical properties of our ScN/GaN layers with thickness $d_\text{ScN}$. Compiled are the total [$n_\text{H}$ ($\mu_\text{H}$)] and the conduction band [$n_\text{CB}$ ($\mu_\text{CB}$)] electron densities (mobilities) at room temperature.}
	\begin{ruledtabular}
		\begin{tabular}{lcccc}
			$d_\text{ScN}$ (nm)    			& 10 				& 15 				& 20 				& 40 \\
			$n_\text{H}$ (cm$^{-3}$)        & 3.1$\x$10$^{19}$  & 1.6$\x$10$^{19}$  & 2.7$\x$10$^{19}$  & 0.7$\x$10$^{19}$ \\
			$n_\text{CB}$ (cm$^{-3}$)        & 1.8$\x$10$^{19}$  & 7.4$\x$10$^{18}$  & 1.2$\x$10$^{19}$  & 3.4$\x$10$^{18}$ \\
			$\mu_\text{H}$ (cm$^{2}$V$^{-1}$s$^{-1}$) & 158     & 48  				& 88  				& 127 \\
			$\mu_\text{CB}$ (cm$^{2}$V$^{-1}$s$^{-1}$) & 204     & 71  				& 130  				& 176 \\
		\end{tabular}
	\end{ruledtabular}
	\label{tab:table}
\end{table}

\section{Results and Discussion}
\subsection{Two-band conduction}
Figure \ref{fig:Hall} shows the temperature dependence of \nHs and \mHs of two ScN(111) layers on GaN:Fe(0001) with different thicknesses. Compared to the ScN(111) reference layer on Al$_2$O$_3$(0001) (see Fig.\,S5 of the supplementary material), the values of \nHs and \mHs are more than one order of magnitude lower and almost two orders of magnitude higher, respectively. Furthermore, both \nHs and \mHs exhibit a characteristic dependence on temperature, in contrast to the temperature-independent behavior observed for the degenerate reference layer. Note that all samples are unambiguously $n$-type for any temperature. We do not see any evidence for a two-dimensional hole gas at the ScN(111)/GaN(0001) interface as recently predicted on theoretical grounds \cite{Adamski2019Dec}.

The comparatively low electron densities suggest a significantly lower concentration of O in the layers on GaN as compared to those on Al$_2$O$_3$. This finding confirms our previous conclusion that the abundance of O in ScN layers on Al$_2$O$_3$ is partly due to out-diffusion from the substrate \cite{Dinh2023Sep}. However, the corresponding reduction in the density of ionized impurities cannot account for the drastically enhanced mobility. Since ScN(111) layers on both GaN(0001) and Al$_2$O$_3$(0001) exhibit a high density of twin boundaries, the factor limiting the mobility for the latter substrate is most likely the very high density of misfit dislocations resulting from the lattice mismatch ($\approx 16\,\%$) for ScN/Al$_2$O$_3$. These misfit dislocations are absent for ScN/GaN due to the virtual lattice match ($\approx 0.1\,\%$).     

\begin{figure}[t]
\includegraphics[width=0.85\columnwidth]{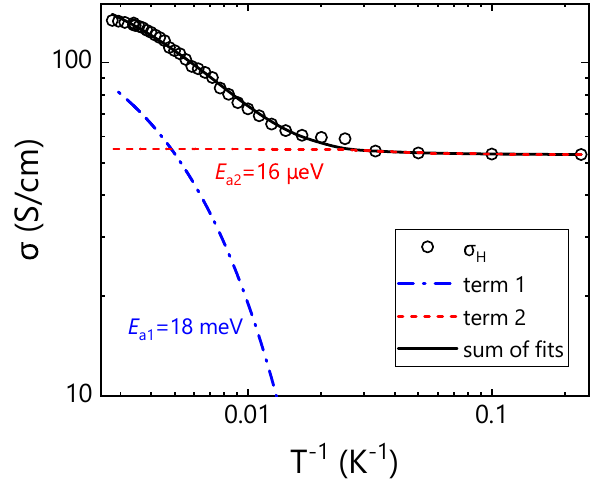}
	\caption{Double-logarithmic Arrhenius representation of the temperature dependence of $\sigma_\text{H}$ for the 15-nm-thick ScN layer (symbols) and a fit using Eq.~\ref{eq:Ea} (lines).}
	\label{fig:Ea}
\end{figure}

Both \nHs and \mHs display the textbook-like behavior of a nondegenerate semiconductor at high temperatures. Specifically, \nHs increases with increasing temperature because of the thermal activation of electrons from shallow donors to the CB \cite{Blakemore1962Janc2}, and \mHs increases and then decreases with increasing temperature due to the combined effects of ionized impurity and phonon scattering \cite{Seeger_1989}. Our data, however, deviate from this behavior at lower temperatures. In particular, \nHs reaches a minimum at a certain temperature $T_\text{IB}$ and then increases again, until it saturates at temperatures $<30$\K. Simultaneously, \mHs saturates at the same temperature. This behavior has been observed in many semiconductors and is indicative of a gradual change from transport in an IB and in the CB at low and high temperatures, respectively \cite{Hung_Phys.Rev._1950,Hung_Phys.Rev._1950a,Hung_Phys.Rev._1954,Morin1954Oct,Fritzsche_Phys.Rev._1955,Conwell1956Jul,Mott_Adv.Phys._1961,Matsubara1961Nov,Blakemore1962Janc3,Kulp1965Oct,Molnar1993Jan,Kabilova2019Feb}. Depending on impurity concentration, the transport in the IB can proceed via hopping with a finite activation energy, or exhibit a metallic character.  

In the presence of two-band conduction, the total conductivity ($\sigma_\text{H}$) can be expressed as the sum of conductivities in the impurity ($\sigma_\text{IB}$) and the conduction bands ($\sigma_\text{CB}$), which are commonly assumed to be thermally activated \cite{Fritzsche1958Jul}:
\begin{align}
	\sigma_\text{H} & = \sigma_\text{CB} + \sigma_\text{IB} \\
	&  = a_{1} e^{-\frac{E_\text{a1}}{k_\text{B} T}} + a_{2} e^{-\frac{E_\text{a2}}{k_\text{B} T}},
	\label{eq:Ea}
\end{align}
where $a_{1}$ and $a_{2}$ are pre-exponential factors, and $E_\text{a1}$ and $E_\text{a2}$ are activation energies. 

Figure~\ref{fig:Ea} shows exemplary conductivity data and a fit with Eq.~\ref{eq:Ea} for the 15-nm-thick ScN layer. The first term ($E_\text{a1} = 18$\,meV) represents the activation of electrons from the IB to the CB, and should be close to the actual ionization energy of the dominant donor (i.\,e., O). In contrast, the second term represents the activation within the IB. In the present case, the fits returns $E_\text{a2} = 16$\,$\muup$eV, close to zero and hence indicating metallic impurity conduction. 

\begin{figure}[b]
	\includegraphics[width=0.85\columnwidth]{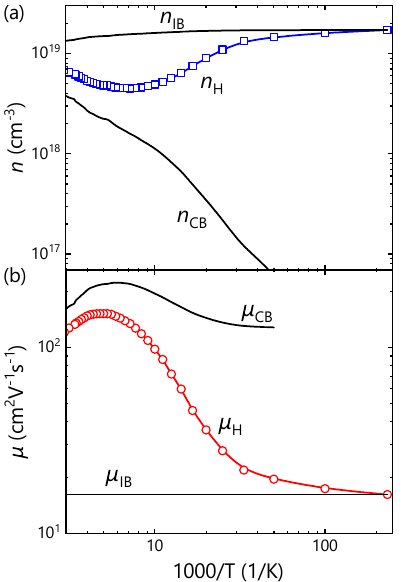}
	\caption{Double-logarithmic Arrhenius representation of \nHs (a) and \mHs (b) for the 40-nm-thick ScN layer. Values of $n_\text{CB, IB}$ in (a) and $\mu_\text{CB, IB}$ in (b) are calculated using Eq.~\ref{eq:2bands}.}
	\label{fig:CB-IB}
\end{figure}

Figure~\ref{fig:CB-IB} shows the Hall data of the 40-nm-thick ScN film. To decompose them into the contributions of the IB and the CB, we use the standard two-band conduction model established by \citet{Hung_Phys.Rev._1950a}:
\begin{subequations}
\begin{align}
  n_\text{H} = \frac{(n_\text{CB} \mu_\text{CB} + n_\text{IB} \mu_\text{IB})^2}{n_\text{CB} \mu_\text{CB}^2 + n_\text{IB} \mu_\text{IB}^2},
  \\
  \mu_\text{H} = \frac{n_\text{CB} \mu_\text{CB}^2 + n_\text{IB} \mu_\text{IB}^2}{n_\text{CB} \mu_\text{CB} + n_\text{IB} \mu_\text{IB}},
\end{align}
\label{eq:2bands}
\end{subequations}
where \nCB, \mCB, \nIBs and \mIBs are the carrier densities and mobilities in the CB and IB, respectively. 

Equations \ref{eq:2bands} contain four unknowns and cannot be solved without additional relations or assumptions. In contrast to two-layer conduction \cite{Petritz1958Jun,Look2008Sep}, for which conduction occurs in parallel but independent channels, the number of carriers in two-band conduction is conserved,
\begin{equation}
n = n_\text{CB} + n_\text{IB} = N_\text{d} - N_\text{a},
\label{Eq:n}
\end{equation}
eliminating one variable, and replacing it with the temperature-independent concentrations of donors ($N_d$) and acceptors ($N_a$). In addition, because of the metallic conduction in the IB, we may assume that \mIBs is constant and equal to $\mu_\text{H}$ at 4\,K. The remaining two unknown variables are \nCB\ and \mCB. Note that this analysis is based on the presumption of a single shallow donor, and a single deep acceptor in ScN. In the case of multiple donor and acceptor species originating, for example, from trace impurities from the Sc source (see Figs.\,S3--S4 in the supplementary material), a quantitative  understanding of the resulting transport is challenging.

\subsection{Electron density in the conduction band}
Values of \nCBs and \mCBs extracted with this procedure are displayed along with the experimental values in Fig.~\ref{fig:CB-IB} for the case of the 40-nm-thick layer. Analogous values extracted for the other layers are shown in Fig.\,S7 of the supplementary material. For distinguishing between nondegenerate and degenerate layers, this figure also shows the effective conduction-band density of states ($N_\text{c}$) given by  
\begin{equation}
	N_\text{c} = \frac{1}{\sqrt{2}} \left(\frac{m^* k_\text{B} T}{\pi \hbar^2}\right)^{3/2}
	\label{eq:Nc}
\end{equation}
with the density-of-states electron mass $m^* = 0.4 m_0$ of ScN \cite{Deng2015Jan,Saha2017Jun}, the Boltzmann constant $k_\text{B}$, and the reduced Planck constant $\hbar$. The data displayed in Fig.\,S7 of the supplementary material show that only the 40-nm-thick layer stays nondegenerate for the whole temperature range, although it approaches degeneracy at temperatures of about 100\,K. In contrast, all the other layers are degenerate at all temperatures. 

Since even the 40-nm-thick layer is close to degeneracy, we use an expression explicitly derived for weakly degenerate semiconductors by \citet{Blakemore1962Janc3}. This expression is valid for a Fermi level not higher than $1.3 k_\text{B} T$ above the CB edge (cf.\ Fig.\,S8 of the supplementary material): 
\begin{widetext}
	\begin{equation}
n_\text{CB} = \frac{2 N_\text{c} (N_\text{d} - N_\text{a})} {\left[N_\text{c} + C (N_\text{d} - N_\text{a}) + \beta^{-1} N_\text{a} e^{E_\text{d}/k_\text{B} T}\right] + \sqrt{\left[N_\text{c} - C (N_\text{d} - N_\text{a}) + \beta^{-1} N_\text{a} e^{E_\text{d}/k_\text{B} T}\right]^2 + 4 \beta^{-1} (N_\text{d} - N_\text{a}) (N_\text{c} + C N_\text{a}) e^{E_\text{d}/k_\text{B} T}}}.
	\label{eq:nCB}
	\end{equation}
\end{widetext}
Here, $C = 0.27$ is a numerical constant that was used to analytically fit the Fermi-Dirac integrals, $\beta = 0.5$ is the degeneracy factor \cite{Blakemore1962Janc3}, and $E_\text{d}$ is the donor ionization energy.

\begin{figure}[b]
	\includegraphics[width=\columnwidth]{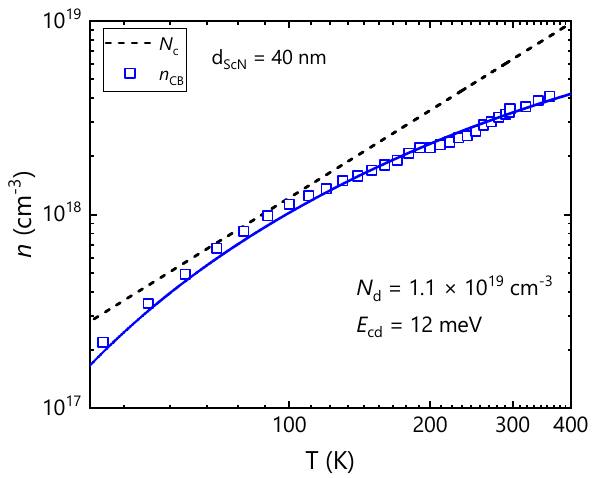}
	\caption{Fit of $n_\text{CB}$ for the 40-nm-thick ScN layer using Eqs.~\ref{eq:Nc}--\ref{eq:nCB}. $N_\text{c}$ is the effective conduction-band density of states of ScN.}
	\label{fig:NdmNa}
\end{figure}

Figure \ref{fig:NdmNa} shows \nCBs for the 40-nm-thick layer as well as a fit of Eqs.~\ref{eq:Nc} and \ref{eq:nCB} to the data to extract $N_\text{d}$, $N_\text{a}$ and $E_\text{d}$. The best fit is obtained with $N_\text{d} = 1 \x 10^{19}$\nunit, $N_\text{a} = 0$, and $E_\text{d} = 12$\,meV, respectively. Fits with finite values for $N_\text{a}$ result in negative values for $E_\text{d}$ and a larger deviation of the fit from the data.

Despite the fact that the fit is unique and in very satisfactory agreement with the data, the quantitative results are probably inaccurate for two reasons. First of all, the total electron density $n = n_\text{H}(4 \,K) = 1.7 \times 10^{19}\,\text{cm}^{-3} > N_d -N_a = 1 \times 10^{19}\,\text{cm}^{-3}$, in contradiction of Eq.~\ref{Eq:n}. Second, in view of the limited purity of the Sc source (see Figs.\,S3--S4 in the supplementary material), we find the complete absence of compensation highly unlikely. Indeed, the 10-nm-thick layer exhibits a higher mobility for simultaneously higher electron density (cf.\ Table~\ref{tab:table} and Fig.~\ref{fig:Hall}), which can only be explained by a considerable degree of compensation for the 40-nm-thick layer. A possible explanation of these discrepancies is the existence of a second, partially filled band formed by deep donors, which is not interacting with the CB, but compensating acceptor states. We did not attempt to quantitatively analyze the data with such a three-band model, since it would add further unknowns to which we have no experimental access.         

The limited thicknesses of the ScN layers under investigation impede secondary ion mass spectrometry (SIMS) for detecting the most abundant impurities. However, XPS measurements performed on cleaned surfaces indicate that O and C are present in the layers [see Fig.\,S3(b) in the supplementary material]. While O is well-known to preferentially incorporate on N sites and to act as a shallow donor in ScN \cite{Dinh2023Sep,Kumagai2018Mar,Deng2015Jan,Smith2001Aug,Saha2017Jun}, the behavior of C in ScN is less well understood. For GaN \cite{Lyons2010Oct}, density functional theory calculations show that C is amphoteric, with the substitution of C atoms on Ga sites (C$_\text{Ga}$) resulting in a shallow donor, while C$_\text{N}$ forms deep acceptor states. For ScN, C$_\text{N}$ has been found be a deep acceptor as well, but the case of C$_\text{Sc}$ has not been investigated \cite{Kumagai2018Mar}. Extrapolating from GaN, it likely that C is amphoteric in ScN as well. Since the Fermi level is close to the CB, the formation energy of C$_\text{N}$ is likely to be lower than that of C$_\text{Sc}$.

Considering the numerous metal impurities present in the Sc source used to grow the ScN layers in the present work (cf.\ Fig.\,S4 in the supplementary material), only a few are present with a concentration that can affect the electrical properties of the layers. With a concentration close to $1 \times 10^{19}$\nunit, Fe is the foremost of these candidates. While Fe is known as a deep acceptor in GaN \cite{Lyons2021Mar}, its behavior in ScN is unknown. Si is dominantly a shallow acceptor in GaN, but may be amphoteric in ScN. Finally, Mg has been found to act as a shallow acceptor in ScN \cite{Kumagai2018Mar}. 

\subsection{Electron mobility and scattering mechanisms}
For many semiconductors, the scattering processes limiting their electron or hole mobility have been investigated in great detail. For materials as diverse as Ge \cite{Debye1954Feb}, GaAs \cite{Fletcher1972Jan}, GaN \cite{Ng1998Aug}, and $\beta$-Ga$_2$O$_3$ \cite{Oishi2015Feb}, ionized-impurity, acoustic and optical phonon scattering are found to determine the mobility at low, medium and high temperatures, respectively. For heteroepitaxial material such as GaN, dislocation scattering may play a role as well \cite{Look1999Feb}. However, for ScN, such an understanding has not been reached since most layers are highly degenerate and thus do not exhibit a clear dependence of mobility on temperature. In the few cases where a quantitative analysis has been possible, the high-temperature slope has been found to follow a power law with an exponent close to that expected for acoustic phonon scattering \cite{Dismukes1972May,Saha2017Jun}. On the other hand, optical phonon scattering is predicted on theoretical grounds to actually dominate scattering at high temperatures \cite{Mu2021Aug}.

(i) To investigate this question in more detail, we first consider a phenomenological expression combining acoustic and ionized impurity phonon scattering \cite{Bikowski2014Oct}: 
\begin{equation}
	\mu_\text{CB} = \left[\left(a T^{-\alpha}\right)^{-1} + \left(b T^{\gamma}\right)^{-1}\right]^{-1}
	\label{eq:acoust}
\end{equation}
with the exponents $\alpha$ and $\gamma$ representing the respective temperature, and the constants $a$ and $b$. For nondegenerate semiconductors, $\alpha = 1.5$ and $\gamma = 1.5$ are theoretically expected and experimentally observed \cite{Debye1954Feb,Fletcher1972Jan,Ng1998Aug,Oishi2015Feb}.

\begin{figure}[t]
	\includegraphics[width=\columnwidth]{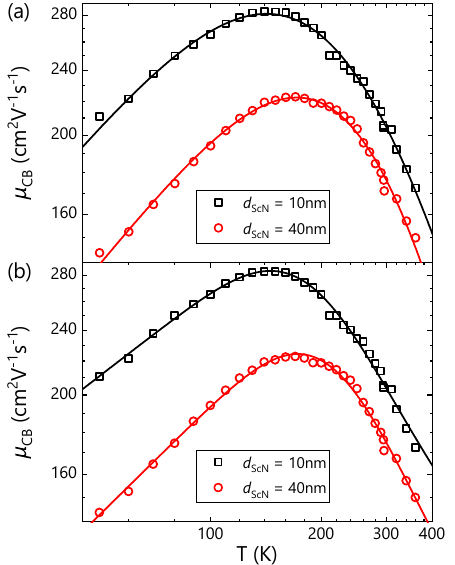}
	\caption{Fits of $\mu_\text{CB}$ of the 10- and 40-nm-thick ScN layers using (a) Eq.~\ref{eq:acoust} and (b) Eq.~\ref{eq:opt}.}
	\label{fig:IIS}
\end{figure}

Figure~\ref{fig:IIS}(a) shows the fit of \mCBs for the 10- and 40-nm-thick layers with Eq.~\ref{eq:acoust}. Evidently, an excellent fit is obtained for both samples, returning values of $\alpha = 1.9 \pm 0.2$ and $\gamma = 0.55 \pm 0.2$. The former of these values is close to the value of 1.85 reported by \citet{Dismukes1972May}. The latter one is smaller than that expected for ionized-impurity scattering in nondegenerate material, since the layers are either close to degeneracy at lower temperatures, or weakly degenerate, for which the temperature dependence of ionized-impurity scattering flattens \cite{Niedermeier_Phys.Rev.B_2017}.  

(ii) Second, we examine the same data by an expression in which the term representing acoustic phonon scattering is replaced by the characteristic temperature dependence for optical phonon scattering in weakly degenerate semiconductors \cite{Bikowski2014Oct}:
\begin{equation}
	\mu_\text{CB} = \left[\frac{1}{c} \left(e^{E_\text{ph}/k_B T}+1\right)^{-1}+\left(d T^{\gamma}\right)^{-1}\right]^{-1},
	\label{eq:opt}
\end{equation}
where $E_\text{ph}$ is the effective phonon energy, and $c$ and $d$ are constants. Fits using Eq.~\ref{eq:opt} are shown in Fig.~\ref{fig:IIS}(b), and are seen to be only marginally different from those displayed in Fig.~\ref{fig:IIS}(a). In fact, the goodness-of-fit as expressed by the adjusted coefficient of determination (adjusted $R^2$) is essentially identical. The effective phonon energy $E_\text{ph} = (61 \pm 5)$\,meV returned from the fits falls within the range of phonon energies typically observed for LO ($\approx 84$\,meV) and TO ($\approx 45$\,meV) phonon modes in ScN (see Raman data of the ScN/GaN layers in Fig.\,S5 of the supplementary material).

The reason for the inability to clearly distinguish between a power-law and an exponential temperature dependence is the comparatively low contrast between the peak mobility at 140--160\,K and the mobility at 360\,K, which in turn is a direct consequence of the dominant ionized-impurity scattering. Note that our layers exhibit rather low electron densities and high mobilities compared to most other reports, except for the one of \citet{Oshima2014Apr}.

\section{Summary and Conclusions}
We have presented an in-depth analysis of the electrical properties of ScN(111) layers grown on nearly lattice-matched GaN:Fe(0001) templates by \PAMBE. Temperature-dependent Hall measurements of these layers, conducted at 4--360\,K in the van der Pauw configuration, have revealed the presence of parallel conduction arising from electrons residing in an impurity and the CB. By employing a two-band conduction model and assuming a single shallow donor in ScN, the electron density and mobility of these bands have been extracted. The 40-nm-thick ScN layer stays nondegenerate throughout the entire temperature range, allowing us to deduce an ionization energy of 12\,meV for the shallow donor. A quantitative analysis of the temperature-dependent mobility reveals that scattering by ionized impurities prevails for temperatures up to 140--180\,K, while phonon scattering takes over at higher temperatures. Fits assuming acoustic or optical phonon scattering are essentially indistinguishable. The latter are, however, theoretically predicted to dominate. In fact, the effective effective phonon energy of $(61 \pm 5)$\,meV returned from the fit assuming optical phonons is just in between the longitudinal optical ($\approx 84$\,meV) and transverse optical ($\approx 45$\,meV) phonon energies in ScN, lending credit to the interpretation supported by theory. Finally, it is worth to note that we did not observe any evidence for a two-dimensional hole gas at the ScN(111)/GaN(0001) interface as recently predicted by modern polarization theory \cite{Adamski2019Dec}. For incoherent interfaces, misfit dislocations form charged line defects that may lead to electrical compensation and scattering, but the interface between ScN and GaN is coherent. We speculate that a high density of positively charged point defects such as N vacancies form at the interface analogously to the case of O vacancies in perovskite heterostructures.

\section*{Acknowledgments}
The semi-insulating GaN:Fe/Al$_2$O$_3$(0001) templates used in this work are courtesy of Stefano Leone from the Fraunhofer-Institut für Angewandte Festkörperphysik. We thank Carsten Stemmler for expert technical assistance with the MBE system and Hua Lv for a critical reading of the manuscript.\\

\noindent
\small{See supplementary material for (1) symmetric $2\theta$--$\omega$ x-ray diffraction (XRD) and pole figure measurements of the ScN\,002 reflection of the 20-nm-thick ScN/GaN layer; (2) high-resolution $2\theta$--$\omega$ XRD scan, corresponding simulation and surface morphology of the 20-nm-thick ScN/GaN layer; (3) XPS spectra of a ScN layer measured before and after Ar$^{+}$ sputtering; (4) Metallic trace contamination of the Sc source supplied by the commercial vendor; (5) Raman spectrum of the 40-nm-thick layer; (6) Temperature-dependent Hall-effect measurements of the 40-nm-thick ScN/GaN:Fe layer and a ScN/Al$_2$O$_3$ reference layer; (7) \nCBs and \mCBs of ScN layers with different thicknesses; (8) $(E_\text{F} - E_\text{c}) / k_\text{B} T$ as a function of temperature calculated for the 40-nm-thick layer.}

\bibliography{BIB_Elektronik_ScN}

\begin{thebibliography}{55}%
\makeatletter
\providecommand \@ifxundefined [1]{%
 \@ifx{#1\undefined}
}%
\providecommand \@ifnum [1]{%
 \ifnum #1\expandafter \@firstoftwo
 \else \expandafter \@secondoftwo
 \fi
}%
\providecommand \@ifx [1]{%
 \ifx #1\expandafter \@firstoftwo
 \else \expandafter \@secondoftwo
 \fi
}%
\providecommand \natexlab [1]{#1}%
\providecommand \enquote  [1]{``#1''}%
\providecommand \bibnamefont  [1]{#1}%
\providecommand \bibfnamefont [1]{#1}%
\providecommand \citenamefont [1]{#1}%
\providecommand \href@noop [0]{\@secondoftwo}%
\providecommand \href [0]{\begingroup \@sanitize@url \@href}%
\providecommand \@href[1]{\@@startlink{#1}\@@href}%
\providecommand \@@href[1]{\endgroup#1\@@endlink}%
\providecommand \@sanitize@url [0]{\catcode `\\12\catcode `\$12\catcode
  `\&12\catcode `\#12\catcode `\^12\catcode `\_12\catcode `\%12\relax}%
\providecommand \@@startlink[1]{}%
\providecommand \@@endlink[0]{}%
\providecommand \url  [0]{\begingroup\@sanitize@url \@url }%
\providecommand \@url [1]{\endgroup\@href {#1}{\urlprefix }}%
\providecommand \urlprefix  [0]{URL }%
\providecommand \Eprint [0]{\href }%
\providecommand \doibase [0]{https://doi.org/}%
\providecommand \selectlanguage [0]{\@gobble}%
\providecommand \bibinfo  [0]{\@secondoftwo}%
\providecommand \bibfield  [0]{\@secondoftwo}%
\providecommand \translation [1]{[#1]}%
\providecommand \BibitemOpen [0]{}%
\providecommand \bibitemStop [0]{}%
\providecommand \bibitemNoStop [0]{.\EOS\space}%
\providecommand \EOS [0]{\spacefactor3000\relax}%
\providecommand \BibitemShut  [1]{\csname bibitem#1\endcsname}%
\let\auto@bib@innerbib\@empty
\bibitem [{\citenamefont {Al-Brithen}\ \emph {et~al.}(2004)\citenamefont
  {Al-Brithen}, \citenamefont {Smith},\ and\ \citenamefont
  {Gall}}]{Al-Brithen2004Jul}%
  \BibitemOpen
  \bibfield  {author} {\bibinfo {author} {\bibfnamefont {H.~A.}\ \bibnamefont
  {Al-Brithen}}, \bibinfo {author} {\bibfnamefont {A.~R.}\ \bibnamefont
  {Smith}},\ and\ \bibinfo {author} {\bibfnamefont {D.}~\bibnamefont {Gall}},\
  }\bibfield  {title} {\bibinfo {title} {{Surface and bulk electronic structure
  of $\mathrm{ScN}(001)$ investigated by scanning tunneling
  microscopy/spectroscopy and optical absorption spectroscopy}},\ }\href
  {https://doi.org/10.1103/PhysRevB.70.045303} {\bibfield  {journal} {\bibinfo
  {journal} {Phys. Rev. B}\ }\textbf {\bibinfo {volume} {70}},\ \bibinfo
  {pages} {045303} (\bibinfo {year} {2004})}\BibitemShut {NoStop}%
\bibitem [{\citenamefont {Saha}\ \emph {et~al.}(2010)\citenamefont {Saha},
  \citenamefont {Acharya}, \citenamefont {Sands},\ and\ \citenamefont
  {Waghmare}}]{Saha2010Feb}%
  \BibitemOpen
  \bibfield  {author} {\bibinfo {author} {\bibfnamefont {B.}~\bibnamefont
  {Saha}}, \bibinfo {author} {\bibfnamefont {J.}~\bibnamefont {Acharya}},
  \bibinfo {author} {\bibfnamefont {T.~D.}\ \bibnamefont {Sands}},\ and\
  \bibinfo {author} {\bibfnamefont {U.~V.}\ \bibnamefont {Waghmare}},\
  }\bibfield  {title} {\bibinfo {title} {{Electronic structure, phonons, and
  thermal properties of ScN, ZrN, and HfN: A first-principles study}},\ }\href
  {https://doi.org/10.1063/1.3291117} {\bibfield  {journal} {\bibinfo
  {journal} {J. Appl. Phys.}\ }\textbf {\bibinfo {volume} {107}},\ \bibinfo
  {pages} {033715} (\bibinfo {year} {2010})}\BibitemShut {NoStop}%
\bibitem [{\citenamefont {Deng}\ \emph {et~al.}(2015)\citenamefont {Deng},
  \citenamefont {Ozsdolay}, \citenamefont {Zheng}, \citenamefont {Khare},\ and\
  \citenamefont {Gall}}]{Deng2015Jan}%
  \BibitemOpen
  \bibfield  {author} {\bibinfo {author} {\bibfnamefont {R.}~\bibnamefont
  {Deng}}, \bibinfo {author} {\bibfnamefont {B.~D.}\ \bibnamefont {Ozsdolay}},
  \bibinfo {author} {\bibfnamefont {P.~Y.}\ \bibnamefont {Zheng}}, \bibinfo
  {author} {\bibfnamefont {S.~V.}\ \bibnamefont {Khare}},\ and\ \bibinfo
  {author} {\bibfnamefont {D.}~\bibnamefont {Gall}},\ }\bibfield  {title}
  {\bibinfo {title} {{Optical and transport measurement and first-principles
  determination of the ScN band gap}},\ }\href
  {https://doi.org/10.1103/PhysRevB.91.045104} {\bibfield  {journal} {\bibinfo
  {journal} {Phys. Rev. B}\ }\textbf {\bibinfo {volume} {91}},\ \bibinfo
  {pages} {045104} (\bibinfo {year} {2015})}\BibitemShut {NoStop}%
\bibitem [{\citenamefont {Cetnar}\ \emph {et~al.}(2018)\citenamefont {Cetnar},
  \citenamefont {Reed}, \citenamefont {Badescu}, \citenamefont {Vangala},
  \citenamefont {Smith},\ and\ \citenamefont {Look}}]{Cetnar2018Nov}%
  \BibitemOpen
  \bibfield  {author} {\bibinfo {author} {\bibfnamefont {J.~S.}\ \bibnamefont
  {Cetnar}}, \bibinfo {author} {\bibfnamefont {A.~N.}\ \bibnamefont {Reed}},
  \bibinfo {author} {\bibfnamefont {S.~C.}\ \bibnamefont {Badescu}}, \bibinfo
  {author} {\bibfnamefont {S.}~\bibnamefont {Vangala}}, \bibinfo {author}
  {\bibfnamefont {H.~A.}\ \bibnamefont {Smith}},\ and\ \bibinfo {author}
  {\bibfnamefont {D.~C.}\ \bibnamefont {Look}},\ }\bibfield  {title} {\bibinfo
  {title} {{Electronic transport in degenerate (100) scandium nitride thin
  films on magnesium oxide substrates}},\ }\href
  {https://doi.org/10.1063/1.5050200} {\bibfield  {journal} {\bibinfo
  {journal} {Appl. Phys. Lett.}\ }\textbf {\bibinfo {volume} {113}},\ \bibinfo
  {pages} {192104} (\bibinfo {year} {2018})}\BibitemShut {NoStop}%
\bibitem [{\citenamefont {Mu}\ \emph {et~al.}(2021)\citenamefont {Mu},
  \citenamefont {Rowberg}, \citenamefont {Leveillee}, \citenamefont
  {Giustino},\ and\ \citenamefont {Van~de Walle}}]{Mu2021Aug}%
  \BibitemOpen
  \bibfield  {author} {\bibinfo {author} {\bibfnamefont {S.}~\bibnamefont
  {Mu}}, \bibinfo {author} {\bibfnamefont {A.~J.~E.}\ \bibnamefont {Rowberg}},
  \bibinfo {author} {\bibfnamefont {J.}~\bibnamefont {Leveillee}}, \bibinfo
  {author} {\bibfnamefont {F.}~\bibnamefont {Giustino}},\ and\ \bibinfo
  {author} {\bibfnamefont {C.~G.}\ \bibnamefont {Van~de Walle}},\ }\bibfield
  {title} {\bibinfo {title} {{First-principles study of electron transport in
  ScN}},\ }\href {https://doi.org/10.1103/PhysRevB.104.075118} {\bibfield
  {journal} {\bibinfo  {journal} {Phys. Rev. B}\ }\textbf {\bibinfo {volume}
  {104}},\ \bibinfo {pages} {075118} (\bibinfo {year} {2021})}\BibitemShut
  {NoStop}%
\bibitem [{\citenamefont {Dismukes}\ \emph {et~al.}(1972)\citenamefont
  {Dismukes}, \citenamefont {Yim},\ and\ \citenamefont
  {Ban}}]{Dismukes1972May}%
  \BibitemOpen
  \bibfield  {author} {\bibinfo {author} {\bibfnamefont {J.~P.}\ \bibnamefont
  {Dismukes}}, \bibinfo {author} {\bibfnamefont {W.~M.}\ \bibnamefont {Yim}},\
  and\ \bibinfo {author} {\bibfnamefont {V.~S.}\ \bibnamefont {Ban}},\
  }\bibfield  {title} {\bibinfo {title} {{Epitaxial growth and properties of
  semiconducting ScN}},\ }\href {https://doi.org/10.1016/0022-0248(72)90185-6}
  {\bibfield  {journal} {\bibinfo  {journal} {J. Cryst. Growth}\ }\textbf
  {\bibinfo {volume} {13-14}},\ \bibinfo {pages} {365} (\bibinfo {year}
  {1972})}\BibitemShut {NoStop}%
\bibitem [{\citenamefont {Smith}\ \emph {et~al.}(2001)\citenamefont {Smith},
  \citenamefont {Al-Brithen}, \citenamefont {Ingram},\ and\ \citenamefont
  {Gall}}]{Smith2001Aug}%
  \BibitemOpen
  \bibfield  {author} {\bibinfo {author} {\bibfnamefont {A.~R.}\ \bibnamefont
  {Smith}}, \bibinfo {author} {\bibfnamefont {H.~A.}\ \bibnamefont
  {Al-Brithen}}, \bibinfo {author} {\bibfnamefont {D.~C.}\ \bibnamefont
  {Ingram}},\ and\ \bibinfo {author} {\bibfnamefont {D.}~\bibnamefont {Gall}},\
  }\bibfield  {title} {\bibinfo {title} {{Molecular beam epitaxy control of the
  structural, optical, and electronic properties of ScN(001)}},\ }\href
  {https://doi.org/10.1063/1.1388161} {\bibfield  {journal} {\bibinfo
  {journal} {J. Appl. Phys.}\ }\textbf {\bibinfo {volume} {90}},\ \bibinfo
  {pages} {1809} (\bibinfo {year} {2001})}\BibitemShut {NoStop}%
\bibitem [{\citenamefont {Moram}\ \emph {et~al.}(2008)\citenamefont {Moram},
  \citenamefont {Barber},\ and\ \citenamefont {Humphreys}}]{Moram2008Oct}%
  \BibitemOpen
  \bibfield  {author} {\bibinfo {author} {\bibfnamefont {M.~A.}\ \bibnamefont
  {Moram}}, \bibinfo {author} {\bibfnamefont {Z.~H.}\ \bibnamefont {Barber}},\
  and\ \bibinfo {author} {\bibfnamefont {C.~J.}\ \bibnamefont {Humphreys}},\
  }\bibfield  {title} {\bibinfo {title} {{The effect of oxygen incorporation in
  sputtered scandium nitride films}},\ }\href
  {https://doi.org/10.1016/j.tsf.2008.05.050} {\bibfield  {journal} {\bibinfo
  {journal} {Thin Solid Films}\ }\textbf {\bibinfo {volume} {516}},\ \bibinfo
  {pages} {8569} (\bibinfo {year} {2008})}\BibitemShut {NoStop}%
\bibitem [{\citenamefont {Saha}\ \emph {et~al.}(2013)\citenamefont {Saha},
  \citenamefont {Naik}, \citenamefont {Drachev}, \citenamefont {Boltasseva},
  \citenamefont {Marinero},\ and\ \citenamefont {Sands}}]{Saha2013Aug}%
  \BibitemOpen
  \bibfield  {author} {\bibinfo {author} {\bibfnamefont {B.}~\bibnamefont
  {Saha}}, \bibinfo {author} {\bibfnamefont {G.}~\bibnamefont {Naik}}, \bibinfo
  {author} {\bibfnamefont {V.~P.}\ \bibnamefont {Drachev}}, \bibinfo {author}
  {\bibfnamefont {A.}~\bibnamefont {Boltasseva}}, \bibinfo {author}
  {\bibfnamefont {E.~E.}\ \bibnamefont {Marinero}},\ and\ \bibinfo {author}
  {\bibfnamefont {T.~D.}\ \bibnamefont {Sands}},\ }\bibfield  {title} {\bibinfo
  {title} {{Electronic and optical properties of ScN and (Sc,Mn)N thin films
  deposited by reactive DC-magnetron sputtering}},\ }\href
  {https://doi.org/10.1063/1.4817715} {\bibfield  {journal} {\bibinfo
  {journal} {J. Appl. Phys.}\ }\textbf {\bibinfo {volume} {114}},\ \bibinfo
  {pages} {063519} (\bibinfo {year} {2013})}\BibitemShut {NoStop}%
\bibitem [{\citenamefont {Oshima}\ \emph {et~al.}(2014)\citenamefont {Oshima},
  \citenamefont {V{\ifmmode\acute{\imath}\else\'{\i}\fi}llora},\ and\
  \citenamefont {Shimamura}}]{Oshima2014Apr}%
  \BibitemOpen
  \bibfield  {author} {\bibinfo {author} {\bibfnamefont {Y.}~\bibnamefont
  {Oshima}}, \bibinfo {author} {\bibfnamefont {E.~G.}\ \bibnamefont
  {V{\ifmmode\acute{\imath}\else\'{\i}\fi}llora}},\ and\ \bibinfo {author}
  {\bibfnamefont {K.}~\bibnamefont {Shimamura}},\ }\bibfield  {title} {\bibinfo
  {title} {{Hydride vapor phase epitaxy and characterization of high-quality
  ScN epilayers}},\ }\href {https://doi.org/10.1063/1.4871656} {\bibfield
  {journal} {\bibinfo  {journal} {J. Appl. Phys.}\ }\textbf {\bibinfo {volume}
  {115}},\ \bibinfo {pages} {153508} (\bibinfo {year} {2014})}\BibitemShut
  {NoStop}%
\bibitem [{\citenamefont {Lupina}\ \emph {et~al.}(2015)\citenamefont {Lupina},
  \citenamefont {Zoellner}, \citenamefont {Niermann}, \citenamefont {Dietrich},
  \citenamefont {Capellini}, \citenamefont {Thapa}, \citenamefont {Haeberlen},
  \citenamefont {Lehmann}, \citenamefont {Storck},\ and\ \citenamefont
  {Schroeder}}]{Lupina2015Nov}%
  \BibitemOpen
  \bibfield  {author} {\bibinfo {author} {\bibfnamefont {L.}~\bibnamefont
  {Lupina}}, \bibinfo {author} {\bibfnamefont {M.~H.}\ \bibnamefont
  {Zoellner}}, \bibinfo {author} {\bibfnamefont {T.}~\bibnamefont {Niermann}},
  \bibinfo {author} {\bibfnamefont {B.}~\bibnamefont {Dietrich}}, \bibinfo
  {author} {\bibfnamefont {G.}~\bibnamefont {Capellini}}, \bibinfo {author}
  {\bibfnamefont {S.~B.}\ \bibnamefont {Thapa}}, \bibinfo {author}
  {\bibfnamefont {M.}~\bibnamefont {Haeberlen}}, \bibinfo {author}
  {\bibfnamefont {M.}~\bibnamefont {Lehmann}}, \bibinfo {author} {\bibfnamefont
  {P.}~\bibnamefont {Storck}},\ and\ \bibinfo {author} {\bibfnamefont
  {T.}~\bibnamefont {Schroeder}},\ }\bibfield  {title} {\bibinfo {title} {{Zero
  lattice mismatch and twin-free single crystalline ScN buffer layers for GaN
  growth on silicon}},\ }\href {https://doi.org/10.1063/1.4935856} {\bibfield
  {journal} {\bibinfo  {journal} {Appl. Phys. Lett.}\ }\textbf {\bibinfo
  {volume} {107}},\ \bibinfo {pages} {201907} (\bibinfo {year}
  {2015})}\BibitemShut {NoStop}%
\bibitem [{\citenamefont {Rao}\ \emph {et~al.}(2020)\citenamefont {Rao},
  \citenamefont {Biswas}, \citenamefont {Flores}, \citenamefont {Chatterjee},
  \citenamefont {Garbrecht}, \citenamefont {Koh}, \citenamefont {Bhatia},
  \citenamefont {Pillai}, \citenamefont {Hopkins}, \citenamefont
  {Martin-Gonzalez},\ and\ \citenamefont {Saha}}]{Rao2020Apr}%
  \BibitemOpen
  \bibfield  {author} {\bibinfo {author} {\bibfnamefont {D.}~\bibnamefont
  {Rao}}, \bibinfo {author} {\bibfnamefont {B.}~\bibnamefont {Biswas}},
  \bibinfo {author} {\bibfnamefont {E.}~\bibnamefont {Flores}}, \bibinfo
  {author} {\bibfnamefont {A.}~\bibnamefont {Chatterjee}}, \bibinfo {author}
  {\bibfnamefont {M.}~\bibnamefont {Garbrecht}}, \bibinfo {author}
  {\bibfnamefont {Y.~R.}\ \bibnamefont {Koh}}, \bibinfo {author} {\bibfnamefont
  {V.}~\bibnamefont {Bhatia}}, \bibinfo {author} {\bibfnamefont {A.~I.~K.}\
  \bibnamefont {Pillai}}, \bibinfo {author} {\bibfnamefont {P.~E.}\
  \bibnamefont {Hopkins}}, \bibinfo {author} {\bibfnamefont {M.}~\bibnamefont
  {Martin-Gonzalez}},\ and\ \bibinfo {author} {\bibfnamefont {B.}~\bibnamefont
  {Saha}},\ }\bibfield  {title} {\bibinfo {title} {{High mobility and high
  thermoelectric power factor in epitaxial ScN thin films deposited with
  plasma-assisted molecular beam epitaxy}},\ }\href
  {https://doi.org/10.1063/5.0004761} {\bibfield  {journal} {\bibinfo
  {journal} {Appl. Phys. Lett.}\ }\textbf {\bibinfo {volume} {116}},\ \bibinfo
  {pages} {152103} (\bibinfo {year} {2020})}\BibitemShut {NoStop}%
\bibitem [{\citenamefont {Dinh}\ \emph
  {et~al.}(2023{\natexlab{a}})\citenamefont {Dinh}, \citenamefont {Peiris},
  \citenamefont {L{\ifmmode\ddot{a}\else\"{a}\fi}hnemann},\ and\ \citenamefont
  {Brandt}}]{Dinh2023Sep}%
  \BibitemOpen
  \bibfield  {author} {\bibinfo {author} {\bibfnamefont {D.~V.}\ \bibnamefont
  {Dinh}}, \bibinfo {author} {\bibfnamefont {F.}~\bibnamefont {Peiris}},
  \bibinfo {author} {\bibfnamefont {J.}~\bibnamefont
  {L{\ifmmode\ddot{a}\else\"{a}\fi}hnemann}},\ and\ \bibinfo {author}
  {\bibfnamefont {O.}~\bibnamefont {Brandt}},\ }\bibfield  {title} {\bibinfo
  {title} {{Optical properties of ScN layers grown on Al$_2$O$_3$(0001) by
  plasma-assisted molecular beam epitaxy}},\ }\href
  {https://doi.org/10.1063/5.0164058} {\bibfield  {journal} {\bibinfo
  {journal} {Appl. Phys. Lett.}\ }\textbf {\bibinfo {volume} {123}},\ \bibinfo
  {pages} {112102} (\bibinfo {year} {2023}{\natexlab{a}})}\BibitemShut
  {NoStop}%
\bibitem [{\citenamefont {Kerdsongpanya}\ \emph {et~al.}(2011)\citenamefont
  {Kerdsongpanya}, \citenamefont {Van~Nong}, \citenamefont {Pryds},
  \citenamefont
  {{\ifmmode\check{Z}\else\v{Z}\fi}ukauskait{\ifmmode\dot{e}\else\.{e}\fi}},
  \citenamefont {Jensen}, \citenamefont {Birch}, \citenamefont {Lu},
  \citenamefont {Hultman}, \citenamefont {Wingqvist},\ and\ \citenamefont
  {Eklund}}]{Kerdsongpanya2011Dec}%
  \BibitemOpen
  \bibfield  {author} {\bibinfo {author} {\bibfnamefont {S.}~\bibnamefont
  {Kerdsongpanya}}, \bibinfo {author} {\bibfnamefont {N.}~\bibnamefont
  {Van~Nong}}, \bibinfo {author} {\bibfnamefont {N.}~\bibnamefont {Pryds}},
  \bibinfo {author} {\bibfnamefont {A.}~\bibnamefont
  {{\ifmmode\check{Z}\else\v{Z}\fi}ukauskait{\ifmmode\dot{e}\else\.{e}\fi}}},
  \bibinfo {author} {\bibfnamefont {J.}~\bibnamefont {Jensen}}, \bibinfo
  {author} {\bibfnamefont {J.}~\bibnamefont {Birch}}, \bibinfo {author}
  {\bibfnamefont {J.}~\bibnamefont {Lu}}, \bibinfo {author} {\bibfnamefont
  {L.}~\bibnamefont {Hultman}}, \bibinfo {author} {\bibfnamefont
  {G.}~\bibnamefont {Wingqvist}},\ and\ \bibinfo {author} {\bibfnamefont
  {P.}~\bibnamefont {Eklund}},\ }\bibfield  {title} {\bibinfo {title}
  {{Anomalously high thermoelectric power factor in epitaxial ScN thin
  films}},\ }\href {https://doi.org/10.1063/1.3665945} {\bibfield  {journal}
  {\bibinfo  {journal} {Appl. Phys. Lett.}\ }\textbf {\bibinfo {volume} {99}},\
  \bibinfo {pages} {232113} (\bibinfo {year} {2011})}\BibitemShut {NoStop}%
\bibitem [{\citenamefont {Burmistrova}\ \emph {et~al.}(2013)\citenamefont
  {Burmistrova}, \citenamefont {Maassen}, \citenamefont {Favaloro},
  \citenamefont {Saha}, \citenamefont {Salamat}, \citenamefont {Rui~Koh},
  \citenamefont {Lundstrom}, \citenamefont {Shakouri},\ and\ \citenamefont
  {Sands}}]{Burmistrova2013Apr}%
  \BibitemOpen
  \bibfield  {author} {\bibinfo {author} {\bibfnamefont {P.~V.}\ \bibnamefont
  {Burmistrova}}, \bibinfo {author} {\bibfnamefont {J.}~\bibnamefont
  {Maassen}}, \bibinfo {author} {\bibfnamefont {T.}~\bibnamefont {Favaloro}},
  \bibinfo {author} {\bibfnamefont {B.}~\bibnamefont {Saha}}, \bibinfo {author}
  {\bibfnamefont {S.}~\bibnamefont {Salamat}}, \bibinfo {author} {\bibfnamefont
  {Y.}~\bibnamefont {Rui~Koh}}, \bibinfo {author} {\bibfnamefont {M.~S.}\
  \bibnamefont {Lundstrom}}, \bibinfo {author} {\bibfnamefont {A.}~\bibnamefont
  {Shakouri}},\ and\ \bibinfo {author} {\bibfnamefont {T.~D.}\ \bibnamefont
  {Sands}},\ }\bibfield  {title} {\bibinfo {title} {{Thermoelectric properties
  of epitaxial ScN films deposited by reactive magnetron sputtering onto
  MgO(001) substrates}},\ }\href {https://doi.org/10.1063/1.4801886} {\bibfield
   {journal} {\bibinfo  {journal} {J. Appl. Phys.}\ }\textbf {\bibinfo {volume}
  {113}},\ \bibinfo {pages} {153704} (\bibinfo {year} {2013})}\BibitemShut
  {NoStop}%
\bibitem [{\citenamefont {Maurya}\ \emph {et~al.}(2022)\citenamefont {Maurya},
  \citenamefont {Rao}, \citenamefont {Acharya}, \citenamefont {Rao},
  \citenamefont {Pillai}, \citenamefont {Selvaraja}, \citenamefont
  {Garbrecht},\ and\ \citenamefont {Saha}}]{Maurya2022Jul}%
  \BibitemOpen
  \bibfield  {author} {\bibinfo {author} {\bibfnamefont {K.~C.}\ \bibnamefont
  {Maurya}}, \bibinfo {author} {\bibfnamefont {D.}~\bibnamefont {Rao}},
  \bibinfo {author} {\bibfnamefont {S.}~\bibnamefont {Acharya}}, \bibinfo
  {author} {\bibfnamefont {P.}~\bibnamefont {Rao}}, \bibinfo {author}
  {\bibfnamefont {A.~I.~K.}\ \bibnamefont {Pillai}}, \bibinfo {author}
  {\bibfnamefont {S.~K.}\ \bibnamefont {Selvaraja}}, \bibinfo {author}
  {\bibfnamefont {M.}~\bibnamefont {Garbrecht}},\ and\ \bibinfo {author}
  {\bibfnamefont {B.}~\bibnamefont {Saha}},\ }\bibfield  {title} {\bibinfo
  {title} {{Polar semiconducting scandium nitride as an infrared plasmon and
  phonon{\textendash}polaritonic material}},\ }\href
  {https://doi.org/10.1021/acs.nanolett.2c00912} {\bibfield  {journal}
  {\bibinfo  {journal} {Nano Lett.}\ }\textbf {\bibinfo {volume} {22}},\
  \bibinfo {pages} {5182} (\bibinfo {year} {2022})}\BibitemShut {NoStop}%
\bibitem [{\citenamefont {Adamski}\ \emph {et~al.}(2019)\citenamefont
  {Adamski}, \citenamefont {Dreyer},\ and\ \citenamefont {Van~de
  Walle}}]{Adamski2019Dec}%
  \BibitemOpen
  \bibfield  {author} {\bibinfo {author} {\bibfnamefont {N.~L.}\ \bibnamefont
  {Adamski}}, \bibinfo {author} {\bibfnamefont {C.~E.}\ \bibnamefont
  {Dreyer}},\ and\ \bibinfo {author} {\bibfnamefont {C.~G.}\ \bibnamefont
  {Van~de Walle}},\ }\bibfield  {title} {\bibinfo {title} {{Giant polarization
  charge density at lattice-matched GaN/ScN interfaces}},\ }\href
  {https://doi.org/10.1063/1.5126717} {\bibfield  {journal} {\bibinfo
  {journal} {Appl. Phys. Lett.}\ }\textbf {\bibinfo {volume} {115}},\ \bibinfo
  {pages} {232103} (\bibinfo {year} {2019})}\BibitemShut {NoStop}%
\bibitem [{\citenamefont {Akiyama}\ \emph {et~al.}(2009)\citenamefont
  {Akiyama}, \citenamefont {Kamohara}, \citenamefont {Kano}, \citenamefont
  {Teshigahara}, \citenamefont {Takeuchi},\ and\ \citenamefont
  {Kawahara}}]{Akiyama2009Feb}%
  \BibitemOpen
  \bibfield  {author} {\bibinfo {author} {\bibfnamefont {M.}~\bibnamefont
  {Akiyama}}, \bibinfo {author} {\bibfnamefont {T.}~\bibnamefont {Kamohara}},
  \bibinfo {author} {\bibfnamefont {K.}~\bibnamefont {Kano}}, \bibinfo {author}
  {\bibfnamefont {A.}~\bibnamefont {Teshigahara}}, \bibinfo {author}
  {\bibfnamefont {Y.}~\bibnamefont {Takeuchi}},\ and\ \bibinfo {author}
  {\bibfnamefont {N.}~\bibnamefont {Kawahara}},\ }\bibfield  {title} {\bibinfo
  {title} {{Enhancement of piezoelectric response in scandium aluminum nitride
  alloy thin films prepared by dual reactive cosputtering}},\ }\href
  {https://doi.org/10.1002/adma.200802611} {\bibfield  {journal} {\bibinfo
  {journal} {Adv. Mater.}\ }\textbf {\bibinfo {volume} {21}},\ \bibinfo {pages}
  {593} (\bibinfo {year} {2009})}\BibitemShut {NoStop}%
\bibitem [{\citenamefont {Hashimoto}\ \emph {et~al.}(2013)\citenamefont
  {Hashimoto}, \citenamefont {Sato}, \citenamefont {Teshigahara}, \citenamefont
  {Nakamura},\ and\ \citenamefont {Kano}}]{Hashimoto2013Mar}%
  \BibitemOpen
  \bibfield  {author} {\bibinfo {author} {\bibfnamefont {K.-y.}\ \bibnamefont
  {Hashimoto}}, \bibinfo {author} {\bibfnamefont {S.}~\bibnamefont {Sato}},
  \bibinfo {author} {\bibfnamefont {A.}~\bibnamefont {Teshigahara}}, \bibinfo
  {author} {\bibfnamefont {T.}~\bibnamefont {Nakamura}},\ and\ \bibinfo
  {author} {\bibfnamefont {K.}~\bibnamefont {Kano}},\ }\bibfield  {title}
  {\bibinfo {title} {{High-performance surface acoustic wave resonators in the
  1 to 3 GHz range using a ScAlN/6H-SiC structure}},\ }\href
  {https://doi.org/10.1109/TUFFC.2013.2606} {\bibfield  {journal} {\bibinfo
  {journal} {IEEE Trans. Ultrason. Ferroelectr. Freq. Control}\ }\textbf
  {\bibinfo {volume} {60}},\ \bibinfo {pages} {637} (\bibinfo {year}
  {2013})}\BibitemShut {NoStop}%
\bibitem [{\citenamefont {Hardy}\ \emph {et~al.}(2017)\citenamefont {Hardy},
  \citenamefont {Downey}, \citenamefont {Nepal}, \citenamefont {Storm},
  \citenamefont {Katzer},\ and\ \citenamefont {Meyer}}]{Hardy2017Apr}%
  \BibitemOpen
  \bibfield  {author} {\bibinfo {author} {\bibfnamefont {M.~T.}\ \bibnamefont
  {Hardy}}, \bibinfo {author} {\bibfnamefont {B.~P.}\ \bibnamefont {Downey}},
  \bibinfo {author} {\bibfnamefont {N.}~\bibnamefont {Nepal}}, \bibinfo
  {author} {\bibfnamefont {D.~F.}\ \bibnamefont {Storm}}, \bibinfo {author}
  {\bibfnamefont {D.~S.}\ \bibnamefont {Katzer}},\ and\ \bibinfo {author}
  {\bibfnamefont {D.~J.}\ \bibnamefont {Meyer}},\ }\bibfield  {title} {\bibinfo
  {title} {{Epitaxial ScAlN grown by molecular beam epitaxy on GaN and SiC
  substrates}},\ }\href {https://doi.org/10.1063/1.4981807} {\bibfield
  {journal} {\bibinfo  {journal} {Appl. Phys. Lett.}\ }\textbf {\bibinfo
  {volume} {110}},\ \bibinfo {pages} {162104} (\bibinfo {year}
  {2017})}\BibitemShut {NoStop}%
\bibitem [{\citenamefont {Wang}\ \emph {et~al.}(2021)\citenamefont {Wang},
  \citenamefont {Wang}, \citenamefont {Wang}, \citenamefont {Mohanty},
  \citenamefont {Diez}, \citenamefont {Wu}, \citenamefont {Sun}, \citenamefont
  {Ahmadi},\ and\ \citenamefont {Mi}}]{Wang2021Aug}%
  \BibitemOpen
  \bibfield  {author} {\bibinfo {author} {\bibfnamefont {P.}~\bibnamefont
  {Wang}}, \bibinfo {author} {\bibfnamefont {D.}~\bibnamefont {Wang}}, \bibinfo
  {author} {\bibfnamefont {B.}~\bibnamefont {Wang}}, \bibinfo {author}
  {\bibfnamefont {S.}~\bibnamefont {Mohanty}}, \bibinfo {author} {\bibfnamefont
  {S.}~\bibnamefont {Diez}}, \bibinfo {author} {\bibfnamefont {Y.}~\bibnamefont
  {Wu}}, \bibinfo {author} {\bibfnamefont {Y.}~\bibnamefont {Sun}}, \bibinfo
  {author} {\bibfnamefont {E.}~\bibnamefont {Ahmadi}},\ and\ \bibinfo {author}
  {\bibfnamefont {Z.}~\bibnamefont {Mi}},\ }\bibfield  {title} {\bibinfo
  {title} {{N-polar ScAlN and HEMTs grown by molecular beam epitaxy}},\ }\href
  {https://doi.org/10.1063/5.0055851} {\bibfield  {journal} {\bibinfo
  {journal} {Appl. Phys. Lett.}\ }\textbf {\bibinfo {volume} {119}},\ \bibinfo
  {pages} {082101} (\bibinfo {year} {2021})}\BibitemShut {NoStop}%
\bibitem [{\citenamefont {Dinh}\ \emph
  {et~al.}(2023{\natexlab{b}})\citenamefont {Dinh}, \citenamefont
  {L{\ifmmode\ddot{a}\else\"{a}\fi}hnemann}, \citenamefont {Geelhaar},\ and\
  \citenamefont {Brandt}}]{Dinh2023Apr}%
  \BibitemOpen
  \bibfield  {author} {\bibinfo {author} {\bibfnamefont {D.~V.}\ \bibnamefont
  {Dinh}}, \bibinfo {author} {\bibfnamefont {J.}~\bibnamefont
  {L{\ifmmode\ddot{a}\else\"{a}\fi}hnemann}}, \bibinfo {author} {\bibfnamefont
  {L.}~\bibnamefont {Geelhaar}},\ and\ \bibinfo {author} {\bibfnamefont
  {O.}~\bibnamefont {Brandt}},\ }\bibfield  {title} {\bibinfo {title} {{Lattice
  parameters of Sc$_x$Al$_{1-x}$N layers grown on GaN(0001) by plasma-assisted
  molecular beam epitaxy}},\ }\href {https://doi.org/10.1063/5.0137873}
  {\bibfield  {journal} {\bibinfo  {journal} {Appl. Phys. Lett.}\ }\textbf
  {\bibinfo {volume} {122}},\ \bibinfo {pages} {152103} (\bibinfo {year}
  {2023}{\natexlab{b}})}\BibitemShut {NoStop}%
\bibitem [{\citenamefont {Wang}\ \emph {et~al.}(2022)\citenamefont {Wang},
  \citenamefont {Wang}, \citenamefont {Mondal},\ and\ \citenamefont
  {Mi}}]{Wang2022Jul}%
  \BibitemOpen
  \bibfield  {author} {\bibinfo {author} {\bibfnamefont {P.}~\bibnamefont
  {Wang}}, \bibinfo {author} {\bibfnamefont {D.}~\bibnamefont {Wang}}, \bibinfo
  {author} {\bibfnamefont {S.}~\bibnamefont {Mondal}},\ and\ \bibinfo {author}
  {\bibfnamefont {Z.}~\bibnamefont {Mi}},\ }\bibfield  {title} {\bibinfo
  {title} {{Ferroelectric N-polar ScAlN/GaN heterostructures grown by molecular
  beam epitaxy}},\ }\href {https://doi.org/10.1063/5.0097117} {\bibfield
  {journal} {\bibinfo  {journal} {Appl. Phys. Lett.}\ }\textbf {\bibinfo
  {volume} {121}},\ \bibinfo {pages} {023501} (\bibinfo {year}
  {2022})}\BibitemShut {NoStop}%
\bibitem [{\citenamefont {Wang}\ \emph {et~al.}(2023)\citenamefont {Wang},
  \citenamefont {Wang}, \citenamefont {Mondal}, \citenamefont {Hu},
  \citenamefont {Wang}, \citenamefont {Wu}, \citenamefont {Ma},\ and\
  \citenamefont {Mi}}]{Wang2023Jan}%
  \BibitemOpen
  \bibfield  {author} {\bibinfo {author} {\bibfnamefont {D.}~\bibnamefont
  {Wang}}, \bibinfo {author} {\bibfnamefont {P.}~\bibnamefont {Wang}}, \bibinfo
  {author} {\bibfnamefont {S.}~\bibnamefont {Mondal}}, \bibinfo {author}
  {\bibfnamefont {M.}~\bibnamefont {Hu}}, \bibinfo {author} {\bibfnamefont
  {D.}~\bibnamefont {Wang}}, \bibinfo {author} {\bibfnamefont {Y.}~\bibnamefont
  {Wu}}, \bibinfo {author} {\bibfnamefont {T.}~\bibnamefont {Ma}},\ and\
  \bibinfo {author} {\bibfnamefont {Z.}~\bibnamefont {Mi}},\ }\bibfield
  {title} {\bibinfo {title} {{Thickness scaling down to 5{\hspace{0.167em}}nm
  of ferroelectric ScAlN on CMOS compatible molybdenum grown by molecular beam
  epitaxy}},\ }\href {https://doi.org/10.1063/5.0136265} {\bibfield  {journal}
  {\bibinfo  {journal} {Appl. Phys. Lett.}\ }\textbf {\bibinfo {volume}
  {122}},\ \bibinfo {pages} {052101} (\bibinfo {year} {2023})}\BibitemShut
  {NoStop}%
\bibitem [{\citenamefont {Casamento}\ \emph {et~al.}(2019)\citenamefont
  {Casamento}, \citenamefont {Wright}, \citenamefont {Chaudhuri}, \citenamefont
  {Xing},\ and\ \citenamefont {Jena}}]{Casamento2019Oct}%
  \BibitemOpen
  \bibfield  {author} {\bibinfo {author} {\bibfnamefont {J.}~\bibnamefont
  {Casamento}}, \bibinfo {author} {\bibfnamefont {J.}~\bibnamefont {Wright}},
  \bibinfo {author} {\bibfnamefont {R.}~\bibnamefont {Chaudhuri}}, \bibinfo
  {author} {\bibfnamefont {H.~G.}\ \bibnamefont {Xing}},\ and\ \bibinfo
  {author} {\bibfnamefont {D.}~\bibnamefont {Jena}},\ }\bibfield  {title}
  {\bibinfo {title} {{Molecular beam epitaxial growth of scandium nitride on
  hexagonal SiC, GaN, and AlN}},\ }\href {https://doi.org/10.1063/1.5121329}
  {\bibfield  {journal} {\bibinfo  {journal} {Appl. Phys. Lett.}\ }\textbf
  {\bibinfo {volume} {115}},\ \bibinfo {pages} {172101} (\bibinfo {year}
  {2019})}\BibitemShut {NoStop}%
\bibitem [{\citenamefont {Ohgaki}\ \emph {et~al.}(2013)\citenamefont {Ohgaki},
  \citenamefont {Watanabe}, \citenamefont {Adachi}, \citenamefont {Sakaguchi},
  \citenamefont {Hishita}, \citenamefont {Ohashi},\ and\ \citenamefont
  {Haneda}}]{Ohgaki2013Sep}%
  \BibitemOpen
  \bibfield  {author} {\bibinfo {author} {\bibfnamefont {T.}~\bibnamefont
  {Ohgaki}}, \bibinfo {author} {\bibfnamefont {K.}~\bibnamefont {Watanabe}},
  \bibinfo {author} {\bibfnamefont {Y.}~\bibnamefont {Adachi}}, \bibinfo
  {author} {\bibfnamefont {I.}~\bibnamefont {Sakaguchi}}, \bibinfo {author}
  {\bibfnamefont {S.}~\bibnamefont {Hishita}}, \bibinfo {author} {\bibfnamefont
  {N.}~\bibnamefont {Ohashi}},\ and\ \bibinfo {author} {\bibfnamefont
  {H.}~\bibnamefont {Haneda}},\ }\bibfield  {title} {\bibinfo {title}
  {{Electrical properties of scandium nitride epitaxial films grown on (100)
  magnesium oxide substrates by molecular beam epitaxy}},\ }\href
  {https://doi.org/10.1063/1.4820391} {\bibfield  {journal} {\bibinfo
  {journal} {J. Appl. Phys.}\ }\textbf {\bibinfo {volume} {114}},\ \bibinfo
  {pages} {093704} (\bibinfo {year} {2013})}\BibitemShut {NoStop}%
\bibitem [{\citenamefont {Al-Atabi}\ \emph {et~al.}(2020)\citenamefont
  {Al-Atabi}, \citenamefont {Zheng}, \citenamefont {Cetnar}, \citenamefont
  {Look}, \citenamefont {Cahill},\ and\ \citenamefont
  {Edgar}}]{Al-Atabi2020Mar}%
  \BibitemOpen
  \bibfield  {author} {\bibinfo {author} {\bibfnamefont {H.}~\bibnamefont
  {Al-Atabi}}, \bibinfo {author} {\bibfnamefont {Q.}~\bibnamefont {Zheng}},
  \bibinfo {author} {\bibfnamefont {J.~S.}\ \bibnamefont {Cetnar}}, \bibinfo
  {author} {\bibfnamefont {D.}~\bibnamefont {Look}}, \bibinfo {author}
  {\bibfnamefont {D.~G.}\ \bibnamefont {Cahill}},\ and\ \bibinfo {author}
  {\bibfnamefont {J.~H.}\ \bibnamefont {Edgar}},\ }\bibfield  {title} {\bibinfo
  {title} {{Properties of bulk scandium nitride crystals grown by physical
  vapor transport}},\ }\href {https://doi.org/10.1063/1.5141808} {\bibfield
  {journal} {\bibinfo  {journal} {Appl. Phys. Lett.}\ }\textbf {\bibinfo
  {volume} {116}},\ \bibinfo {pages} {132103} (\bibinfo {year}
  {2020})}\BibitemShut {NoStop}%
\bibitem [{\citenamefont {Saha}\ \emph {et~al.}(2017)\citenamefont {Saha},
  \citenamefont {Garbrecht}, \citenamefont {Perez-Taborda}, \citenamefont
  {Fawey}, \citenamefont {Koh}, \citenamefont {Shakouri}, \citenamefont
  {Martin-Gonzalez}, \citenamefont {Hultman},\ and\ \citenamefont
  {Sands}}]{Saha2017Jun}%
  \BibitemOpen
  \bibfield  {author} {\bibinfo {author} {\bibfnamefont {B.}~\bibnamefont
  {Saha}}, \bibinfo {author} {\bibfnamefont {M.}~\bibnamefont {Garbrecht}},
  \bibinfo {author} {\bibfnamefont {J.~A.}\ \bibnamefont {Perez-Taborda}},
  \bibinfo {author} {\bibfnamefont {M.~H.}\ \bibnamefont {Fawey}}, \bibinfo
  {author} {\bibfnamefont {Y.~R.}\ \bibnamefont {Koh}}, \bibinfo {author}
  {\bibfnamefont {A.}~\bibnamefont {Shakouri}}, \bibinfo {author}
  {\bibfnamefont {M.}~\bibnamefont {Martin-Gonzalez}}, \bibinfo {author}
  {\bibfnamefont {L.}~\bibnamefont {Hultman}},\ and\ \bibinfo {author}
  {\bibfnamefont {T.~D.}\ \bibnamefont {Sands}},\ }\bibfield  {title} {\bibinfo
  {title} {{Compensation of native donor doping in ScN: Carrier concentration
  control and p-type ScN}},\ }\href {https://doi.org/10.1063/1.4989530}
  {\bibfield  {journal} {\bibinfo  {journal} {Appl. Phys. Lett.}\ }\textbf
  {\bibinfo {volume} {110}},\ \bibinfo {pages} {252104} (\bibinfo {year}
  {2017})}\BibitemShut {NoStop}%
\bibitem [{\citenamefont {Blakemore}(1962{\natexlab{a}})}]{Blakemore1962Janc2}%
  \BibitemOpen
  \bibfield  {author} {\bibinfo {author} {\bibfnamefont {J.~S.}\ \bibnamefont
  {Blakemore}},\ }\bibfield  {title} {\bibinfo {title} {{Chapter 2 - The Fermi
  level - electron density equilibrium}},\ }in\ \href
  {https://doi.org/10.1016/B978-0-08-009592-9.50006-6} {\emph {\bibinfo
  {booktitle} {{Semiconductor statistics}}}}\ (\bibinfo  {publisher}
  {Pergamon},\ \bibinfo {address} {Oxford, England, UK},\ \bibinfo {year}
  {1962})\ pp.\ \bibinfo {pages} {75--116}\BibitemShut {NoStop}%
\bibitem [{\citenamefont {Seeger}(1989)}]{Seeger_1989}%
  \BibitemOpen
  \bibfield  {author} {\bibinfo {author} {\bibfnamefont {K.}~\bibnamefont
  {Seeger}},\ }\href {https://doi.org/10.1007/978-3-662-02576-5} {\emph
  {\bibinfo {title} {Semiconductor Physics}}},\ \bibinfo {edition} {4th}\ ed.,\
  edited by\ \bibinfo {editor} {\bibfnamefont {M.}~\bibnamefont {Cardona}},
  \bibinfo {editor} {\bibfnamefont {P.}~\bibnamefont {Fulde}}, \bibinfo
  {editor} {\bibfnamefont {K.}~\bibnamefont {{von Klitzing}}}, \bibinfo
  {editor} {\bibfnamefont {H.-J.}\ \bibnamefont {Queisser}},\ and\ \bibinfo
  {editor} {\bibfnamefont {H.~K.~V.}\ \bibnamefont {Lotsch}},\ \bibinfo
  {series} {Springer series in Solid-State Sciences}, Vol.~\bibinfo {volume}
  {40}\ (\bibinfo  {publisher} {{Springer}},\ \bibinfo {address} {{Berlin
  Heidelberg}},\ \bibinfo {year} {1989})\BibitemShut {NoStop}%
\bibitem [{\citenamefont {Hung}\ and\ \citenamefont
  {Gliessman}(1950)}]{Hung_Phys.Rev._1950}%
  \BibitemOpen
  \bibfield  {author} {\bibinfo {author} {\bibfnamefont {C.~S.}\ \bibnamefont
  {Hung}}\ and\ \bibinfo {author} {\bibfnamefont {J.~R.}\ \bibnamefont
  {Gliessman}},\ }\bibfield  {title} {\bibinfo {title} {The resistivity and
  {{Hall}} effect of germanium at low temperatures},\ }\href
  {https://doi.org/10.1103/PhysRev.79.726} {\bibfield  {journal} {\bibinfo
  {journal} {Phys. Rev.}\ }\textbf {\bibinfo {volume} {79}},\ \bibinfo {pages}
  {726} (\bibinfo {year} {1950})}\BibitemShut {NoStop}%
\bibitem [{\citenamefont {Hung}(1950)}]{Hung_Phys.Rev._1950a}%
  \BibitemOpen
  \bibfield  {author} {\bibinfo {author} {\bibfnamefont {C.~S.}\ \bibnamefont
  {Hung}},\ }\bibfield  {title} {\bibinfo {title} {Theory of resistivity and
  {{Hall}} effect at very low temperatures},\ }\href
  {https://doi.org/10.1103/PhysRev.79.727} {\bibfield  {journal} {\bibinfo
  {journal} {Phys. Rev.}\ }\textbf {\bibinfo {volume} {79}},\ \bibinfo {pages}
  {727} (\bibinfo {year} {1950})}\BibitemShut {NoStop}%
\bibitem [{\citenamefont {Hung}\ and\ \citenamefont
  {Gliessman}(1954)}]{Hung_Phys.Rev._1954}%
  \BibitemOpen
  \bibfield  {author} {\bibinfo {author} {\bibfnamefont {C.~S.}\ \bibnamefont
  {Hung}}\ and\ \bibinfo {author} {\bibfnamefont {J.~R.}\ \bibnamefont
  {Gliessman}},\ }\bibfield  {title} {\bibinfo {title} {Resistivity and
  {{Hall}} effect of germanium at low temperatures},\ }\href
  {https://doi.org/10.1103/PhysRev.96.1226} {\bibfield  {journal} {\bibinfo
  {journal} {Phys. Rev.}\ }\textbf {\bibinfo {volume} {96}},\ \bibinfo {pages}
  {1226} (\bibinfo {year} {1954})}\BibitemShut {NoStop}%
\bibitem [{\citenamefont {Morin}\ and\ \citenamefont
  {Maita}(1954)}]{Morin1954Oct}%
  \BibitemOpen
  \bibfield  {author} {\bibinfo {author} {\bibfnamefont {F.~J.}\ \bibnamefont
  {Morin}}\ and\ \bibinfo {author} {\bibfnamefont {J.~P.}\ \bibnamefont
  {Maita}},\ }\bibfield  {title} {\bibinfo {title} {{Electrical properties of
  silicon containing arsenic and boron}},\ }\href
  {https://doi.org/10.1103/PhysRev.96.28} {\bibfield  {journal} {\bibinfo
  {journal} {Phys. Rev.}\ }\textbf {\bibinfo {volume} {96}},\ \bibinfo {pages}
  {28} (\bibinfo {year} {1954})}\BibitemShut {NoStop}%
\bibitem [{\citenamefont {Fritzsche}(1955)}]{Fritzsche_Phys.Rev._1955}%
  \BibitemOpen
  \bibfield  {author} {\bibinfo {author} {\bibfnamefont {H.}~\bibnamefont
  {Fritzsche}},\ }\bibfield  {title} {\bibinfo {title} {Electrical properties
  of germanium semiconductors at low temperatures},\ }\href
  {https://doi.org/10.1103/PhysRev.99.406} {\bibfield  {journal} {\bibinfo
  {journal} {Phys. Rev.}\ }\textbf {\bibinfo {volume} {99}},\ \bibinfo {pages}
  {406} (\bibinfo {year} {1955})}\BibitemShut {NoStop}%
\bibitem [{\citenamefont {Conwell}(1956)}]{Conwell1956Jul}%
  \BibitemOpen
  \bibfield  {author} {\bibinfo {author} {\bibfnamefont {E.~M.}\ \bibnamefont
  {Conwell}},\ }\bibfield  {title} {\bibinfo {title} {{Impurity band conduction
  in germanium and silicon}},\ }\href {https://doi.org/10.1103/PhysRev.103.51}
  {\bibfield  {journal} {\bibinfo  {journal} {Phys. Rev.}\ }\textbf {\bibinfo
  {volume} {103}},\ \bibinfo {pages} {51} (\bibinfo {year} {1956})}\BibitemShut
  {NoStop}%
\bibitem [{\citenamefont {Mott}\ and\ \citenamefont
  {Twose}(1961)}]{Mott_Adv.Phys._1961}%
  \BibitemOpen
  \bibfield  {author} {\bibinfo {author} {\bibfnamefont {N.}~\bibnamefont
  {Mott}}\ and\ \bibinfo {author} {\bibfnamefont {W.}~\bibnamefont {Twose}},\
  }\bibfield  {title} {\bibinfo {title} {The theory of impurity conduction},\
  }\href {https://doi.org/10.1080/00018736100101271} {\bibfield  {journal}
  {\bibinfo  {journal} {Advances in Physics}\ }\textbf {\bibinfo {volume}
  {10}},\ \bibinfo {pages} {107} (\bibinfo {year} {1961})}\BibitemShut
  {NoStop}%
\bibitem [{\citenamefont {Matsubara}\ and\ \citenamefont
  {Toyozawa}(1961)}]{Matsubara1961Nov}%
  \BibitemOpen
  \bibfield  {author} {\bibinfo {author} {\bibfnamefont {T.}~\bibnamefont
  {Matsubara}}\ and\ \bibinfo {author} {\bibfnamefont {Y.}~\bibnamefont
  {Toyozawa}},\ }\bibfield  {title} {\bibinfo {title} {{Theory of impurity band
  conduction in semiconductors: An approach to random lattice problem}},\
  }\href {https://doi.org/10.1143/PTP.26.739} {\bibfield  {journal} {\bibinfo
  {journal} {Prog. Theor. Phys.}\ }\textbf {\bibinfo {volume} {26}},\ \bibinfo
  {pages} {739} (\bibinfo {year} {1961})}\BibitemShut {NoStop}%
\bibitem [{\citenamefont {Blakemore}(1962{\natexlab{b}})}]{Blakemore1962Janc3}%
  \BibitemOpen
  \bibfield  {author} {\bibinfo {author} {\bibfnamefont {J.~S.}\ \bibnamefont
  {Blakemore}},\ }\bibfield  {title} {\bibinfo {title} {{Chapter 3 -
  Semiconductors dominated by impurity levels}},\ }in\ \href
  {https://doi.org/10.1016/B978-0-08-009592-9.50007-8} {\emph {\bibinfo
  {booktitle} {{Semiconductor statistics}}}}\ (\bibinfo  {publisher}
  {Pergamon},\ \bibinfo {address} {Oxford, England, UK},\ \bibinfo {year}
  {1962})\ pp.\ \bibinfo {pages} {117--176}\BibitemShut {NoStop}%
\bibitem [{\citenamefont {Kulp}\ \emph {et~al.}(1965)\citenamefont {Kulp},
  \citenamefont {Gale},\ and\ \citenamefont {Schulze}}]{Kulp1965Oct}%
  \BibitemOpen
  \bibfield  {author} {\bibinfo {author} {\bibfnamefont {B.~A.}\ \bibnamefont
  {Kulp}}, \bibinfo {author} {\bibfnamefont {K.~A.}\ \bibnamefont {Gale}},\
  and\ \bibinfo {author} {\bibfnamefont {R.~G.}\ \bibnamefont {Schulze}},\
  }\bibfield  {title} {\bibinfo {title} {{Impurity conductivity in
  single-crystal CdS}},\ }\href {https://doi.org/10.1103/PhysRev.140.A252}
  {\bibfield  {journal} {\bibinfo  {journal} {Phys. Rev.}\ }\textbf {\bibinfo
  {volume} {140}},\ \bibinfo {pages} {A252} (\bibinfo {year}
  {1965})}\BibitemShut {NoStop}%
\bibitem [{\citenamefont {Molnar}\ \emph {et~al.}(1993)\citenamefont {Molnar},
  \citenamefont {Lei},\ and\ \citenamefont {Moustakas}}]{Molnar1993Jan}%
  \BibitemOpen
  \bibfield  {author} {\bibinfo {author} {\bibfnamefont {R.~J.}\ \bibnamefont
  {Molnar}}, \bibinfo {author} {\bibfnamefont {T.}~\bibnamefont {Lei}},\ and\
  \bibinfo {author} {\bibfnamefont {T.~D.}\ \bibnamefont {Moustakas}},\
  }\bibfield  {title} {\bibinfo {title} {{Electron transport mechanism in
  gallium nitride}},\ }\href {https://doi.org/10.1063/1.108823} {\bibfield
  {journal} {\bibinfo  {journal} {Appl. Phys. Lett.}\ }\textbf {\bibinfo
  {volume} {62}},\ \bibinfo {pages} {72} (\bibinfo {year} {1993})}\BibitemShut
  {NoStop}%
\bibitem [{\citenamefont {Kabilova}\ \emph {et~al.}(2019)\citenamefont
  {Kabilova}, \citenamefont {Kurdak},\ and\ \citenamefont
  {Peterson}}]{Kabilova2019Feb}%
  \BibitemOpen
  \bibfield  {author} {\bibinfo {author} {\bibfnamefont {Z.}~\bibnamefont
  {Kabilova}}, \bibinfo {author} {\bibfnamefont {C.}~\bibnamefont {Kurdak}},\
  and\ \bibinfo {author} {\bibfnamefont {R.~L.}\ \bibnamefont {Peterson}},\
  }\bibfield  {title} {\bibinfo {title} {{Observation of impurity band
  conduction and variable range hopping in heavily doped (010)
  {$\beta$}-Ga$_2$O$_3$}},\ }\href {https://doi.org/10.1088/1361-6641/ab0150}
  {\bibfield  {journal} {\bibinfo  {journal} {Semicond. Sci. Technol.}\
  }\textbf {\bibinfo {volume} {34}},\ \bibinfo {pages} {03LT02} (\bibinfo
  {year} {2019})}\BibitemShut {NoStop}%
\bibitem [{\citenamefont {Fritzsche}(1958)}]{Fritzsche1958Jul}%
  \BibitemOpen
  \bibfield  {author} {\bibinfo {author} {\bibfnamefont {H.}~\bibnamefont
  {Fritzsche}},\ }\bibfield  {title} {\bibinfo {title} {{Resistivity and
  {{Hall}} coefficient of antimony-doped germanium at low temperatures}},\
  }\href {https://doi.org/10.1016/0022-3697(58)90220-8} {\bibfield  {journal}
  {\bibinfo  {journal} {J. Phys. Chem. Solids}\ }\textbf {\bibinfo {volume}
  {6}},\ \bibinfo {pages} {69} (\bibinfo {year} {1958})}\BibitemShut {NoStop}%
\bibitem [{\citenamefont {Petritz}(1958)}]{Petritz1958Jun}%
  \BibitemOpen
  \bibfield  {author} {\bibinfo {author} {\bibfnamefont {R.~L.}\ \bibnamefont
  {Petritz}},\ }\bibfield  {title} {\bibinfo {title} {{Theory of an experiment
  for measuring the mobility and density of carriers in the space-charge region
  of a semiconductor surface}},\ }\href
  {https://doi.org/10.1103/PhysRev.110.1254} {\bibfield  {journal} {\bibinfo
  {journal} {Phys. Rev.}\ }\textbf {\bibinfo {volume} {110}},\ \bibinfo {pages}
  {1254} (\bibinfo {year} {1958})}\BibitemShut {NoStop}%
\bibitem [{\citenamefont {Look}(2008)}]{Look2008Sep}%
  \BibitemOpen
  \bibfield  {author} {\bibinfo {author} {\bibfnamefont {D.~C.}\ \bibnamefont
  {Look}},\ }\bibfield  {title} {\bibinfo {title} {{Two-layer Hall-effect model
  with arbitrary surface-donor profiles: application to ZnO}},\ }\bibfield
  {journal} {\bibinfo  {journal} {J. Appl. Phys.}\ }\textbf {\bibinfo {volume}
  {104}},\ \href {https://doi.org/10.1063/1.2986143} {10.1063/1.2986143}
  (\bibinfo {year} {2008})\BibitemShut {NoStop}%
\bibitem [{\citenamefont {Kumagai}\ \emph {et~al.}(2018)\citenamefont
  {Kumagai}, \citenamefont {Tsunoda},\ and\ \citenamefont
  {Oba}}]{Kumagai2018Mar}%
  \BibitemOpen
  \bibfield  {author} {\bibinfo {author} {\bibfnamefont {Y.}~\bibnamefont
  {Kumagai}}, \bibinfo {author} {\bibfnamefont {N.}~\bibnamefont {Tsunoda}},\
  and\ \bibinfo {author} {\bibfnamefont {F.}~\bibnamefont {Oba}},\ }\bibfield
  {title} {\bibinfo {title} {{Point defects and $p$-type doping in ScN from
  first principles}},\ }\href {https://doi.org/10.1103/PhysRevApplied.9.034019}
  {\bibfield  {journal} {\bibinfo  {journal} {Phys. Rev. Appl.}\ }\textbf
  {\bibinfo {volume} {9}},\ \bibinfo {pages} {034019} (\bibinfo {year}
  {2018})}\BibitemShut {NoStop}%
\bibitem [{\citenamefont {Lyons}\ \emph {et~al.}(2010)\citenamefont {Lyons},
  \citenamefont {Janotti},\ and\ \citenamefont {Van~de Walle}}]{Lyons2010Oct}%
  \BibitemOpen
  \bibfield  {author} {\bibinfo {author} {\bibfnamefont {J.~L.}\ \bibnamefont
  {Lyons}}, \bibinfo {author} {\bibfnamefont {A.}~\bibnamefont {Janotti}},\
  and\ \bibinfo {author} {\bibfnamefont {C.~G.}\ \bibnamefont {Van~de Walle}},\
  }\bibfield  {title} {\bibinfo {title} {{Carbon impurities and the yellow
  luminescence in GaN}},\ }\href {https://doi.org/10.1063/1.3492841} {\bibfield
   {journal} {\bibinfo  {journal} {Appl. Phys. Lett.}\ }\textbf {\bibinfo
  {volume} {97}},\ \bibinfo {pages} {152108} (\bibinfo {year}
  {2010})}\BibitemShut {NoStop}%
\bibitem [{\citenamefont {Lyons}\ \emph {et~al.}(2021)\citenamefont {Lyons},
  \citenamefont {Wickramaratne},\ and\ \citenamefont {Van~de
  Walle}}]{Lyons2021Mar}%
  \BibitemOpen
  \bibfield  {author} {\bibinfo {author} {\bibfnamefont {J.~L.}\ \bibnamefont
  {Lyons}}, \bibinfo {author} {\bibfnamefont {D.}~\bibnamefont
  {Wickramaratne}},\ and\ \bibinfo {author} {\bibfnamefont {C.~G.}\
  \bibnamefont {Van~de Walle}},\ }\bibfield  {title} {\bibinfo {title} {{A
  first-principles understanding of point defects and impurities in GaN}},\
  }\href {https://doi.org/10.1063/5.0041506} {\bibfield  {journal} {\bibinfo
  {journal} {J. Appl. Phys.}\ }\textbf {\bibinfo {volume} {129}},\ \bibinfo
  {pages} {111101} (\bibinfo {year} {2021})}\BibitemShut {NoStop}%
\bibitem [{\citenamefont {Debye}\ and\ \citenamefont
  {Conwell}(1954)}]{Debye1954Feb}%
  \BibitemOpen
  \bibfield  {author} {\bibinfo {author} {\bibfnamefont {P.~P.}\ \bibnamefont
  {Debye}}\ and\ \bibinfo {author} {\bibfnamefont {E.~M.}\ \bibnamefont
  {Conwell}},\ }\bibfield  {title} {\bibinfo {title} {{Electrical properties of
  $n$-type Germanium}},\ }\href {https://doi.org/10.1103/PhysRev.93.693}
  {\bibfield  {journal} {\bibinfo  {journal} {Phys. Rev.}\ }\textbf {\bibinfo
  {volume} {93}},\ \bibinfo {pages} {693} (\bibinfo {year} {1954})}\BibitemShut
  {NoStop}%
\bibitem [{\citenamefont {Fletcher}\ and\ \citenamefont
  {Butcher}(1972)}]{Fletcher1972Jan}%
  \BibitemOpen
  \bibfield  {author} {\bibinfo {author} {\bibfnamefont {K.}~\bibnamefont
  {Fletcher}}\ and\ \bibinfo {author} {\bibfnamefont {P.~N.}\ \bibnamefont
  {Butcher}},\ }\bibfield  {title} {\bibinfo {title} {{An exact solution of the
  linearized Boltzmann equation with applications to the Hall mobility and Hall
  factor of n-GaAs}},\ }\href {https://doi.org/10.1088/0022-3719/5/2/010}
  {\bibfield  {journal} {\bibinfo  {journal} {J. Phys. C: Solid State Phys.}\
  }\textbf {\bibinfo {volume} {5}},\ \bibinfo {pages} {212} (\bibinfo {year}
  {1972})}\BibitemShut {NoStop}%
\bibitem [{\citenamefont {Ng}\ \emph {et~al.}(1998)\citenamefont {Ng},
  \citenamefont {Doppalapudi}, \citenamefont {Moustakas}, \citenamefont
  {Weimann},\ and\ \citenamefont {Eastman}}]{Ng1998Aug}%
  \BibitemOpen
  \bibfield  {author} {\bibinfo {author} {\bibfnamefont {H.~M.}\ \bibnamefont
  {Ng}}, \bibinfo {author} {\bibfnamefont {D.}~\bibnamefont {Doppalapudi}},
  \bibinfo {author} {\bibfnamefont {T.~D.}\ \bibnamefont {Moustakas}}, \bibinfo
  {author} {\bibfnamefont {N.~G.}\ \bibnamefont {Weimann}},\ and\ \bibinfo
  {author} {\bibfnamefont {L.~F.}\ \bibnamefont {Eastman}},\ }\bibfield
  {title} {\bibinfo {title} {{The role of dislocation scattering in n-type GaN
  films}},\ }\href {https://doi.org/10.1063/1.122012} {\bibfield  {journal}
  {\bibinfo  {journal} {Appl. Phys. Lett.}\ }\textbf {\bibinfo {volume} {73}},\
  \bibinfo {pages} {821} (\bibinfo {year} {1998})}\BibitemShut {NoStop}%
\bibitem [{\citenamefont {Oishi}\ \emph {et~al.}(2015)\citenamefont {Oishi},
  \citenamefont {Koga}, \citenamefont {Harada},\ and\ \citenamefont
  {Kasu}}]{Oishi2015Feb}%
  \BibitemOpen
  \bibfield  {author} {\bibinfo {author} {\bibfnamefont {T.}~\bibnamefont
  {Oishi}}, \bibinfo {author} {\bibfnamefont {Y.}~\bibnamefont {Koga}},
  \bibinfo {author} {\bibfnamefont {K.}~\bibnamefont {Harada}},\ and\ \bibinfo
  {author} {\bibfnamefont {M.}~\bibnamefont {Kasu}},\ }\bibfield  {title}
  {\bibinfo {title} {{High-mobility {$\beta$}-Ga$_2$O$_3(\bar{2}01)$ single
  crystals grown by edge-defined film-fed growth method and their Schottky
  barrier diodes with Ni contact}},\ }\href
  {https://doi.org/10.7567/APEX.8.031101} {\bibfield  {journal} {\bibinfo
  {journal} {Appl. Phys. Express}\ }\textbf {\bibinfo {volume} {8}},\ \bibinfo
  {pages} {031101} (\bibinfo {year} {2015})}\BibitemShut {NoStop}%
\bibitem [{\citenamefont {Look}\ and\ \citenamefont
  {Sizelove}(1999)}]{Look1999Feb}%
  \BibitemOpen
  \bibfield  {author} {\bibinfo {author} {\bibfnamefont {D.~C.}\ \bibnamefont
  {Look}}\ and\ \bibinfo {author} {\bibfnamefont {J.~R.}\ \bibnamefont
  {Sizelove}},\ }\bibfield  {title} {\bibinfo {title} {{Dislocation scattering
  in GaN}},\ }\href {https://doi.org/10.1103/PhysRevLett.82.1237} {\bibfield
  {journal} {\bibinfo  {journal} {Phys. Rev. Lett.}\ }\textbf {\bibinfo
  {volume} {82}},\ \bibinfo {pages} {1237} (\bibinfo {year}
  {1999})}\BibitemShut {NoStop}%
\bibitem [{\citenamefont {Bikowski}\ and\ \citenamefont
  {Ellmer}(2014)}]{Bikowski2014Oct}%
  \BibitemOpen
  \bibfield  {author} {\bibinfo {author} {\bibfnamefont {A.}~\bibnamefont
  {Bikowski}}\ and\ \bibinfo {author} {\bibfnamefont {K.}~\bibnamefont
  {Ellmer}},\ }\bibfield  {title} {\bibinfo {title} {{Analytical model of
  electron transport in polycrystalline, degenerately doped ZnO films}},\
  }\href {https://doi.org/10.1063/1.4896839} {\bibfield  {journal} {\bibinfo
  {journal} {J. Appl. Phys.}\ }\textbf {\bibinfo {volume} {116}},\ \bibinfo
  {pages} {143704} (\bibinfo {year} {2014})}\BibitemShut {NoStop}%
\bibitem [{\citenamefont {Niedermeier}\ \emph {et~al.}(2017)\citenamefont
  {Niedermeier}, \citenamefont {Rhode}, \citenamefont {Ide}, \citenamefont
  {Hiramatsu}, \citenamefont {Hosono}, \citenamefont {Kamiya},\ and\
  \citenamefont {Moram}}]{Niedermeier_Phys.Rev.B_2017}%
  \BibitemOpen
  \bibfield  {author} {\bibinfo {author} {\bibfnamefont {C.~A.}\ \bibnamefont
  {Niedermeier}}, \bibinfo {author} {\bibfnamefont {S.}~\bibnamefont {Rhode}},
  \bibinfo {author} {\bibfnamefont {K.}~\bibnamefont {Ide}}, \bibinfo {author}
  {\bibfnamefont {H.}~\bibnamefont {Hiramatsu}}, \bibinfo {author}
  {\bibfnamefont {H.}~\bibnamefont {Hosono}}, \bibinfo {author} {\bibfnamefont
  {T.}~\bibnamefont {Kamiya}},\ and\ \bibinfo {author} {\bibfnamefont {M.~A.}\
  \bibnamefont {Moram}},\ }\bibfield  {title} {\bibinfo {title} {Electron
  effective mass and mobility limits in degenerate perovskite stannate
  {BaSnO$_3$}},\ }\href {https://doi.org/10.1103/PhysRevB.95.161202} {\bibfield
   {journal} {\bibinfo  {journal} {Phys. Rev. B}\ }\textbf {\bibinfo {volume}
  {95}},\ \bibinfo {pages} {161202} (\bibinfo {year} {2017})}\BibitemShut
  {NoStop}%
\end{thebibliography}%


\begin{thebibliography}{4}%
\makeatletter
\providecommand \@ifxundefined [1]{%
 \@ifx{#1\undefined}
}%
\providecommand \@ifnum [1]{%
 \ifnum #1\expandafter \@firstoftwo
 \else \expandafter \@secondoftwo
 \fi
}%
\providecommand \@ifx [1]{%
 \ifx #1\expandafter \@firstoftwo
 \else \expandafter \@secondoftwo
 \fi
}%
\providecommand \natexlab [1]{#1}%
\providecommand \enquote  [1]{``#1''}%
\providecommand \bibnamefont  [1]{#1}%
\providecommand \bibfnamefont [1]{#1}%
\providecommand \citenamefont [1]{#1}%
\providecommand \href@noop [0]{\@secondoftwo}%
\providecommand \href [0]{\begingroup \@sanitize@url \@href}%
\providecommand \@href[1]{\@@startlink{#1}\@@href}%
\providecommand \@@href[1]{\endgroup#1\@@endlink}%
\providecommand \@sanitize@url [0]{\catcode `\\12\catcode `\$12\catcode
  `\&12\catcode `\#12\catcode `\^12\catcode `\_12\catcode `\%12\relax}%
\providecommand \@@startlink[1]{}%
\providecommand \@@endlink[0]{}%
\providecommand \url  [0]{\begingroup\@sanitize@url \@url }%
\providecommand \@url [1]{\endgroup\@href {#1}{\urlprefix }}%
\providecommand \urlprefix  [0]{URL }%
\providecommand \Eprint [0]{\href }%
\providecommand \doibase [0]{https://doi.org/}%
\providecommand \selectlanguage [0]{\@gobble}%
\providecommand \bibinfo  [0]{\@secondoftwo}%
\providecommand \bibfield  [0]{\@secondoftwo}%
\providecommand \translation [1]{[#1]}%
\providecommand \BibitemOpen [0]{}%
\providecommand \bibitemStop [0]{}%
\providecommand \bibitemNoStop [0]{.\EOS\space}%
\providecommand \EOS [0]{\spacefactor3000\relax}%
\providecommand \BibitemShut  [1]{\csname bibitem#1\endcsname}%
\let\auto@bib@innerbib\@empty
\bibitem [{\citenamefont {Lichtensteiger}(2018)}]{Lichtensteiger}%
  \BibitemOpen
  \bibfield  {author} {\bibinfo {author} {\bibfnamefont {C.}~\bibnamefont
  {Lichtensteiger}},\ }\bibfield  {title} {\bibinfo {title}
  {{InteractiveXRDFit: a new tool to simulate and fit X-ray diffractograms of
  oxide thin films and heterostructures}},\ }\href
  {https://doi.org/10.1107/S1600576718012840} {\bibfield  {journal} {\bibinfo
  {journal} {J. Appl. Crystallogr.}\ }\textbf {\bibinfo {volume} {51}},\
  \bibinfo {pages} {1745} (\bibinfo {year} {2018})}\BibitemShut {NoStop}%
\bibitem [{\citenamefont {Blakemore}(1962)}]{Blakemore1962Janc3}%
  \BibitemOpen
  \bibfield  {author} {\bibinfo {author} {\bibfnamefont {J.~S.}\ \bibnamefont
  {Blakemore}},\ }\bibfield  {title} {\bibinfo {title} {{Chapter 3 -
  Semiconductors dominated by impurity levels}},\ }in\ \href
  {https://doi.org/10.1016/B978-0-08-009592-9.50007-8} {\emph {\bibinfo
  {booktitle} {{Semiconductor statistics}}}}\ (\bibinfo  {publisher}
  {Pergamon},\ \bibinfo {address} {Oxford, England, UK},\ \bibinfo {year}
  {1962})\ pp.\ \bibinfo {pages} {117--176}\BibitemShut {NoStop}%
\bibitem [{\citenamefont {Deng}\ \emph {et~al.}(2015)\citenamefont {Deng},
  \citenamefont {Ozsdolay}, \citenamefont {Zheng}, \citenamefont {Khare},\ and\
  \citenamefont {Gall}}]{Deng2015Jan}%
  \BibitemOpen
  \bibfield  {author} {\bibinfo {author} {\bibfnamefont {R.}~\bibnamefont
  {Deng}}, \bibinfo {author} {\bibfnamefont {B.~D.}\ \bibnamefont {Ozsdolay}},
  \bibinfo {author} {\bibfnamefont {P.~Y.}\ \bibnamefont {Zheng}}, \bibinfo
  {author} {\bibfnamefont {S.~V.}\ \bibnamefont {Khare}},\ and\ \bibinfo
  {author} {\bibfnamefont {D.}~\bibnamefont {Gall}},\ }\bibfield  {title}
  {\bibinfo {title} {{Optical and transport measurement and first-principles
  determination of the ScN band gap}},\ }\href
  {https://doi.org/10.1103/PhysRevB.91.045104} {\bibfield  {journal} {\bibinfo
  {journal} {Phys. Rev. B}\ }\textbf {\bibinfo {volume} {91}},\ \bibinfo
  {pages} {045104} (\bibinfo {year} {2015})}\BibitemShut {NoStop}%
\bibitem [{\citenamefont {Saha}\ \emph {et~al.}(2017)\citenamefont {Saha},
  \citenamefont {Garbrecht}, \citenamefont {Perez-Taborda}, \citenamefont
  {Fawey}, \citenamefont {Koh}, \citenamefont {Shakouri}, \citenamefont
  {Martin-Gonzalez}, \citenamefont {Hultman},\ and\ \citenamefont
  {Sands}}]{Saha2017Jun}%
  \BibitemOpen
  \bibfield  {author} {\bibinfo {author} {\bibfnamefont {B.}~\bibnamefont
  {Saha}}, \bibinfo {author} {\bibfnamefont {M.}~\bibnamefont {Garbrecht}},
  \bibinfo {author} {\bibfnamefont {J.~A.}\ \bibnamefont {Perez-Taborda}},
  \bibinfo {author} {\bibfnamefont {M.~H.}\ \bibnamefont {Fawey}}, \bibinfo
  {author} {\bibfnamefont {Y.~R.}\ \bibnamefont {Koh}}, \bibinfo {author}
  {\bibfnamefont {A.}~\bibnamefont {Shakouri}}, \bibinfo {author}
  {\bibfnamefont {M.}~\bibnamefont {Martin-Gonzalez}}, \bibinfo {author}
  {\bibfnamefont {L.}~\bibnamefont {Hultman}},\ and\ \bibinfo {author}
  {\bibfnamefont {T.~D.}\ \bibnamefont {Sands}},\ }\bibfield  {title} {\bibinfo
  {title} {{Compensation of native donor doping in ScN: Carrier concentration
  control and p-type ScN}},\ }\href {https://doi.org/10.1063/1.4989530}
  {\bibfield  {journal} {\bibinfo  {journal} {Appl. Phys. Lett.}\ }\textbf
  {\bibinfo {volume} {110}},\ \bibinfo {pages} {252104} (\bibinfo {year}
  {2017})}\BibitemShut {NoStop}%
\end{thebibliography}%

\end{document}


	
	\title{Electrical properties of ScN(111) layers grown on GaN(0001)\\ by plasma-assisted molecular beam epitaxy: supplementary material}
	
	\author{Duc V. Dinh}
	\email[Electronic email: ]{dinh@pdi-berlin.de}
	\author{Oliver Brandt}
	\affiliation{Paul-Drude-Institut für Festkörperelektronik, Leibniz-Institut im Forschungsverbund Berlin e.V., Hausvogteiplatz 5--7, 10117 Berlin, Germany.}

\maketitle

\begin{figure}[t]
	\includegraphics[width=0.875\columnwidth]{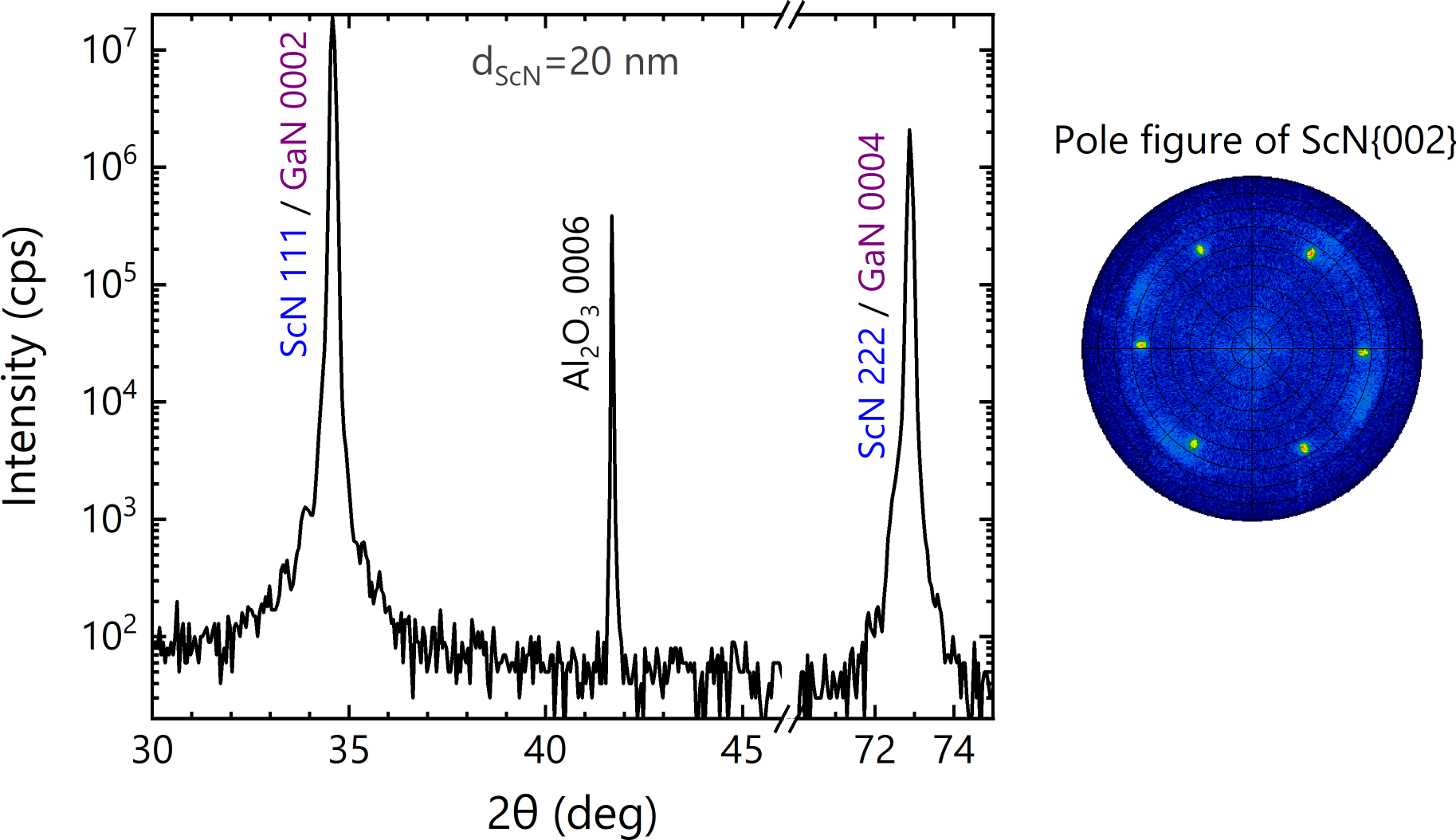}
	\caption{Symmetric $2\theta-\omega$ x-ray diffraction scan (performed with open detector, Philips PANalytical X’Pert PRO MRD) of the 20-nm-thick ScN(111) layer grown on a nearly lattice-matched Fe-doped GaN/Al$_2$O$_3$(0001) template. Thickness fringes around the ScN\,111 reflection indicate the sharp interface between ScN/GaN. Pole figure measurement of the ScN\{002\} reflection (2$\theta = 40.0$°) of this layer shows six peaks at $\psi \approx 54.7$°, indicating the twinned growth of the ScN layer on GaN(0001).}
\end{figure}

\begin{figure}[t]
	\includegraphics[width=0.65\columnwidth]{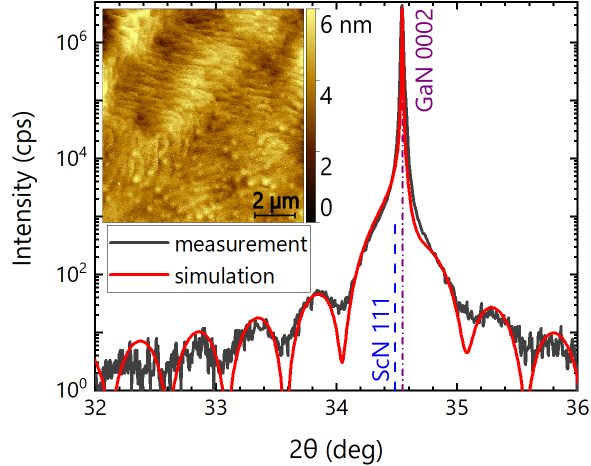}
	\caption{Symmetric $2\theta-\omega$ x-ray diffraction scan (performed with analyzer) of the 20-nm-thick ScN(111) layer grown on GaN(0001) template. The simulation has been performed using InteractiveXRDFit \cite{Lichtensteiger}. The dash-dotted and dashed lines indicate the 2$\theta$ position of the 111 and 0002 reflections of relaxed ScN(111) and GaN(0001), respectively. The \textit{inset} shows a $10 \x 10$\,$\muup$m$^2$ atomic force topograph (Dimension Edge, Bruker) of this layer with a root-mean-square roughness value of 0.8\,nm.}
\end{figure}

\begin{figure}
	\centering
	\includegraphics[width=0.55\linewidth]{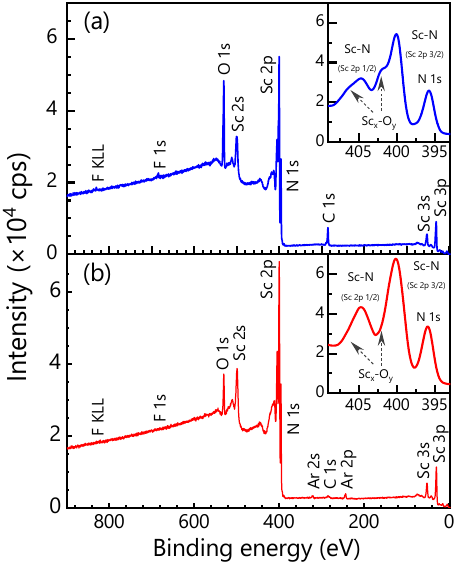}
	\caption{XPS survey spectra measured (a) before and (b) after Ar$^{+}$ sputtering (Scienta Omicron). Note the change in intensities of O\,1\textit{s}, Sc peaks, and C\,1\textit{s} in (b). The insets show high-resolution spectra scanned around Sc\,2\textit{p} and N\,1\textit{s}.}
\end{figure}

\begin{figure}
	\centering
	\includegraphics[width=\linewidth]{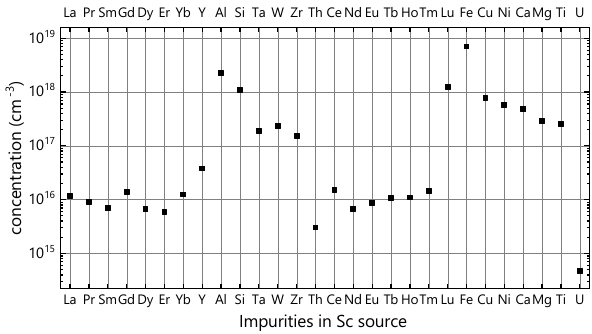}
	\caption{Metal and silicon impurity data in the Sc source (99.999\% pure) as supplied by the commercial vendor.}
\end{figure}

\begin{figure}
	\centering
	\includegraphics[width=0.7\linewidth]{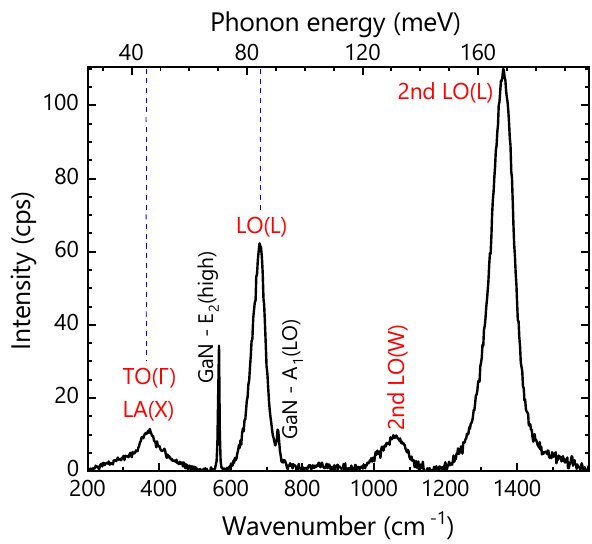}
	\caption{Raman spectrum of the 40-nm-thick layer measured with a 473-nm diode-pumped solid state laser (Horiba LabRAM HR Evolution). The phonon peaks E$_2$(high) and A$_1$(LO) of GaN are at 567\,cm$^{-1}$ and 730\,cm$^{-1}$, respectively.}
\end{figure}

\begin{figure}
	\centering
	\includegraphics[width=0.8\linewidth]{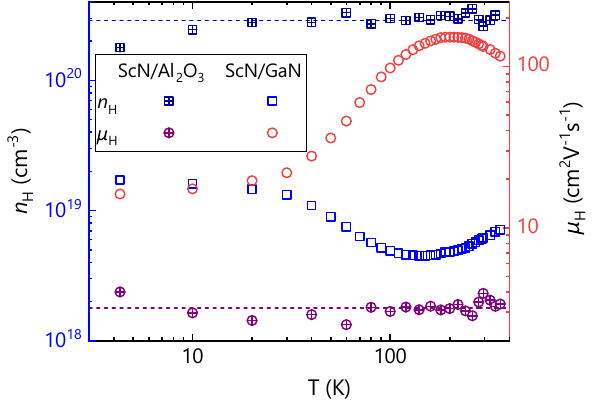}
	\caption{Temperature-dependent Hall-effect measurements of the 40-nm-thick ScN/GaN:Fe layer and a ScN/Al$_2$O$_3$ reference layer.}
\end{figure}

\begin{figure}
	\centering
	\includegraphics[width=0.75\linewidth]{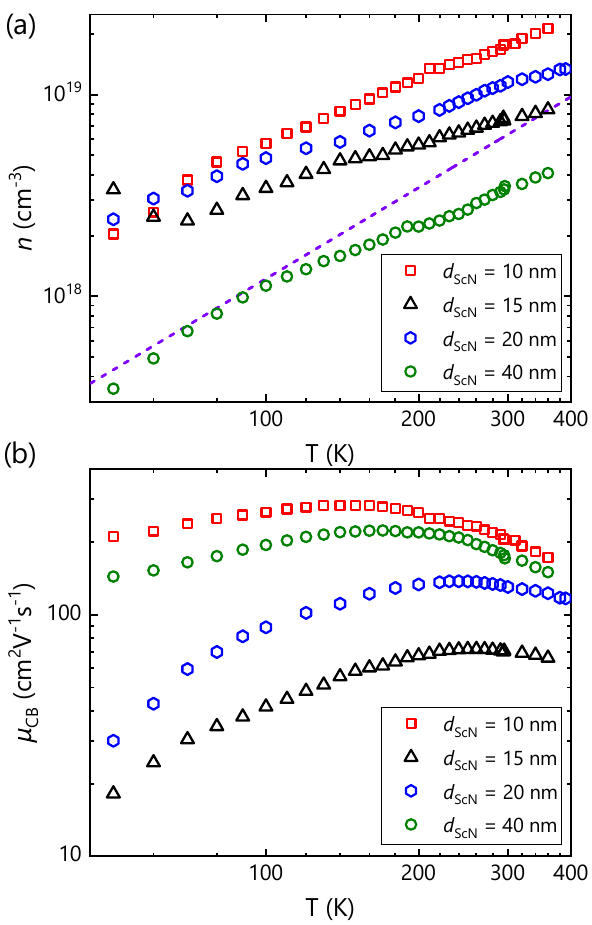}
	\caption{(a) $n_\text{CB}$ and (b) $\mu_\text{CB}$ of the ScN layers grown with different thicknesses. The dashed-line in (a) indicates the effective conduction-band density of states $N_\text{c}$ of ScN.}
\end{figure}

\clearpage

\begin{figure}
	\centering
	\includegraphics[width=0.6\linewidth]{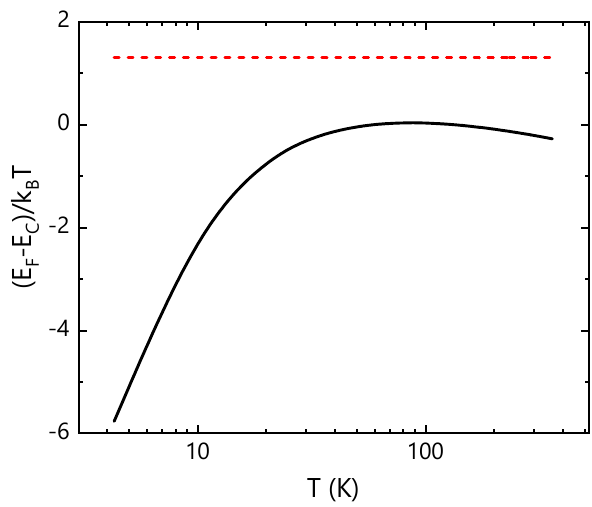}
	\caption{$(E_\text{F} - E_\text{c}) / k_\text{B} T$ as a function of temperature $T$ calculated for the 40-nm-thick layer.}
	\label{fig:OK}
\end{figure}

The expression used in the main text to analyze the temperature dependence of the electrons density in the conduction band is valid for a Fermi level not higher than $1.3 k_\text{B} T$ above the conduction band edge. The position of the Fermi level  $E_\text{F}$ is given by \cite{Blakemore1962Janc3}:

\begin{equation}
\label{eq:Ef}
e^{\frac{E_\text{F} - E_\text{c}}{k_\text{B} T}}  = \frac{2 (N_\text{d} - N_\text{a})} {\left(N_\text{c} - C (N_\text{d} - N_\text{a}) + \beta^{-1} N_\text{a} e^{\frac{E_\text{d}}{k_\text{B} T}}\right) + \sqrt{\left(N_\text{c} - C (N_\text{d} - N_\text{a}) + \beta^{-1} N_\text{a} e^{\frac{E_\text{d}}{k_\text{B} T}}\right)^2 + 4 \beta^{-1} (N_\text{d} - N_\text{a}) (N_\text{c} + C N_\text{a}) e^{\frac{E_\text{d}}{k_\text{B} T}}}},
\end{equation}

where $C = 0.27$ is a numerical constant that was used to analytically fit the Fermi-Dirac integrals, $\beta=0.5$ is the degeneracy factor, $k_\text{B}$ is the Boltzmann constant. $E_\text{d}$ is the donor ionization energy. $N_\text{c}$ is the effective density of states in the conduction band given by:
\begin{equation}
	N_\text{c} = \frac{1}{\sqrt{2}} \left(\frac{m^* k_\text{B} T}{\pi \hbar^2}\right)^{3/2}
	\label{eq:Nc}
\end{equation}
with the density-of-states electron mass $m^* = 0.4 m_0$ of ScN \cite{Deng2015Jan,Saha2017Jun}, the Boltzmann constant $k_\text{B}$, and the reduced Planck constant $\hbar$.

Figure~\ref{fig:OK} shows that with $N_\text{d} = 1 \times 10^{19}$\,cm$^{-3}$ and $E_\text{cd} = 12$\,meV calculated for the 40-nm-thick layer, Eq.~6 in the main text is valid for the whole temperature range.\\\\

\bibliography{BIB_Elektronik_ScN}